\documentclass[a4paper,11pt]{article}

\usepackage{jcappub}

\usepackage{slashed}
\usepackage{youngtab}

\newcommand\fverb{\setbox\fverbbox=\hbox\bgroup\verb}
\newcommand\fverbdo{\egroup\medskip\noindent%
            \fbox{\unhbox\fverbbox}\ }
\newcommand\fverbit{\egroup\item[\fbox{\unhbox\fverbbox}]}
\newbox\fverbbox
\newcommand{\nablaslash}{\not{\hbox{\kern-3pt $\nabla$}}}

\title{Cosmological three-coupled scalar theory for the dS/LCFT correspondence}

\author{Yun~Soo~Myung}
\author{and~Taeyoon~Moon}
\affiliation{Institute of Basic Science and Department of Computer
Simulation, Inje University,\\
Gimhae 621-749, Korea}

\emailAdd{ysmyung@inje.ac.kr} \emailAdd{tymoon@inje.ac.kr}

\abstract{We investigate  cosmological perturbations generated
during de Sitter inflation in the three-coupled scalar theory. This
theory is composed of three coupled scalars ($\phi_p,p=1,2,3$) to
give a sixth-order derivative  scalar theory for $\phi_3$, in
addition to tensor. Recovering the power spectra between scalars
from the LCFT correlators in momentum space indicates that the de
Sitter/logarithmic conformal field theory (dS/LCFT) correspondence
works  in the superhorizon limit. We use LCFT correlators derived
from the dS/LCFT differentiate dictionary to compare cosmological
correlators (power spectra) and find also LCFT correlators by making
use of extrapolate dictionary. This is because the former approach
is more conventional than the latter.  A bulk version dual to the
truncation process to find a unitary CFT in the LCFT corresponds to
selecting a physical field $\phi_2$ with positive norm propagating
on the dS spacetime.}

\begin{document}

\maketitle \flushbottom
\section{Introduction}
The Lee-Wick (LW) model of a fourth-order derivative scalar theory
with $\hat{\phi}$ has provided a cosmological bounce which could
avoid the big bang singularity~\cite{Cai:2008qw}. By introducing an
auxiliary field (LW scalar $\tilde{\phi}$) and   redefining the
normal scalar field as $\phi=\hat{\phi}+\tilde{\phi}$, the
fourth-order Lagrangian can be expressed in terms of two
second-order Lagrangians where the kinetic and  mass terms of the LW
scalar have the opposite sign compared the signs for the normal
scalar. The LW scalar plays the role of a ghost scalar and thus, it
is responsible for giving a bouncing solution. In the contracting
phase, $\phi$ dominates while $\phi$ freezes and $\tilde{\phi}$
still oscillates near the bounce. In the expanding phase, $\phi$
dominates again.  Vacuum fluctuations in the contracting phase have
led to a scale-invariant spectrum of cosmological perturbations.
Recently, it was proposed that the bounce inflation scenario can
simultaneously explain the Planck and BICEP2 observations better
than the $\Lambda$CDM model~\cite{Xia:2014tda}.

On the other hand,  the singleton theory~\cite{Flato:1986uh}  was
used widely to derive the anti de Sitter (AdS)/logarithmic conformal
field theory (LCFT)
correspondence~\cite{Ghezelbash:1998rj,Kogan:1999bn,Myung:1999nd,Grumiller:2013at}
as well as  the dS/LCFT
correspondence~\cite{Kehagias:2012pd,Myung:2014pza}. The singleton
  is a bulk Lagrangian composed of dipole fields ($\phi_1,\phi_2$)
to give a fourth-order differential equation
$(\nabla^2-m^2)^2\phi_2=0$ for $\phi_2$ [equivalently, a
second-order coupled equation $(\nabla^2-m^2)\phi_2=\mu^2\phi_1$] on
the AdS/dS background, even though its starting Lagrangian  is
second-order. The field $\phi_1$ can be seen as an auxiliary field
to lower the number of derivatives in the fourth-order Lagrangian.
The similarity between the LW model and singleton theory is the
connection of
($\phi\leftrightarrow\phi_1,\hat{\phi}\leftrightarrow\phi_2$) and a
difference is the absence of the LW scalar $\tilde{\phi}$ in the
singleton theory. Also, the LW model has two different masses, but
the singleton has the same mass. Here we are interested in studying
the singleton in the AdS/dS background  induced by the
negative/positive cosmological constants.  To that end, the
singleton was used to derive the
LCFT~\cite{Gurarie:1993xq,Flohr:2001zs} on the boundary of AdS/dS
which induces a non-unitraity problem.  The singleton was of
interest from the cosmological point of view for two reasons: The
power spectrum during dS inflation is not scale-invariant even in
the limit of zero-mass because of the presence of
$(H/2\pi)^2\ln[\epsilon k]$ in the superhorizon and it provides one
example in which the Suyama-Yamaguchi inequality is
reversed~\cite{Kehagias:2012pd}. This inequality describes the
connection between the collapsed limit of  four-point correlator and
the squeezed limit of  three-point correlator~\cite{Suyama:2007bg}.

In order to resolve the non-unitarity problem confronted in the
singleton, one has to truncate log-modes out by imposing the
appropriate AdS boundary conditions~\cite{Bergshoeff:2012sc}. A rank
of the LCFT refers to the dimensionality of the Jordan cell. The
rank-2 LCFT dual to a critical gravity has a rank-2 Jordan cell and
thus, an operator has a logarithmic (log) partner. Stating simply, a
log partner is dual to $\phi_2$ whose equation is a fourth-order
equation. However, there remains nothing for the rank-2 LCFT after
making truncation. This implies that there is no power spectrum. The
LCFT dual to a tricritical gravity has a rank-3 Jordan
cell~\cite{Bergshoeff:2012ev} and  an operator has two log partners
of log and log$^2$.  After truncation, there remains a unitary
subspace with non-negative state. For simplicity, it is natural to
consider a sixth-order scalar field theory to cast off the
non-unitarity problem. In order to avoid a difficulty in dealing
with a single sixth-order theory, we introduce an equivalent
three-coupled scalar fields ($\phi_1,\phi_2,\phi_3$) with degenerate
masses $m^2$~\cite{Kim:2013waf,Kim:2013mfa}. A bulk version dual to
the truncation process to find a unitary CFT in the rank-3 LCFT
corresponds to selecting a physical field $\phi_2$ with positive
norm propagating on the dS spacetime. After truncation, the only
non-zero power spectrum will be ${\cal P}_{22,0}=\xi^2(\epsilon
k)^{2w}$ which is surly non-negative.

 At this stage, we would like to mention that this three-coupled
scalar theory might be  considered as a toy model of the tricritical
gravity. However, it was pointed out that these linearized
approaches of tricritical gravities have pathologies when
considering the non-linear level~\cite{Apolo:2012vv}. This implies
that calculations on the linearized level seemed to lend support to
the possibility of truncating the theory. In this sense, we have to
regard our model of the three-coupled scalar  as a toy model of
(linearized) tricritical gravities.

The canonical quantization of three-coupled scalar theory was
performed with nontrivial commutation relations  on the Minkowski
spacetime. These commutation relations will be used  to compute the
power spectra of scalars when one chooses the Bunch-Davies (BD)
vacuum in the subhorizon limit of dS inflation. This is considered
as the dS/quantum field theory (QFT) correspondence in the
subhorizon limit.

A single scalar field (inflaton) with a canonical kinetic term is
generally  known to be a promising model for describing the
slow-roll (dS-like) inflation~\cite{Baumann:2008aq,Baumann:2009ds}.
Importantly, a recent detection of B-mode polarization  has enhanced
the occurrence  of inflation at the GUT scale~\cite{Ade:2014xna}.
Also, it is worth noting that  the dS/CFT
correspondence~\cite{Maldacena:2002vr} has firstly provided the
derivation of the non-Gaussianity from the single field inflation in
the superhorizon limit.~\cite{Chen:2006nt}.  If one accepts
holographic inflation such that the dS inflation era of our universe
is  described by a dual CFT living on the slice ($\mathbb{R}^3$) at
the end of inflation, the BICEP2 results might determine the central
charge of the CFT~\cite{Larsen:2014wpa}.

Accordingly, it is promising to compute the power spectrum of the
three-coupled scalar field theory generated during dS inflation
because this theory provides non-canonical  fourth-order and
sixth-order equations, in addition to the canonical second-order
equation.  In order to compute the power spectrum, one needs to
choose the BD vacuum in the subhorizon limit of $z\to \infty$ (UV
region). Here, one has to quantize the three-coupled scalar fields
canonically in the subhorizon limit whose commutation relations
(\ref{scft}) are important to compute the power spectrum. This may
provide a hint for the dS/QFT correspondence in the subhorizon
region of $z\to \infty$ (UV region).  Also, it is meaningful to
check whether the dS/LCFT correspondence plays a crucial role in
computing the power spectrum in the superhorizon limit of $z\to 0$
(IR region)~\cite{Kehagias:2012pd}.  We will observe that the
commutation relations (\ref{scft}) between  three scalars have the
similar form as LCFT correlators (\ref{lcft-mat}), which shows the
double correspondences of dS/QFT and dS/LCFT
 on the UV and IR regions, respectively.

\section{Three-coupled scalar field theory }

We consider  the three-coupled scalar field theory where three
fields ($\phi_1,\phi_2,\phi_3$) are coupled minimally to Einstein
gravity as~\cite{Bergshoeff:2012sc,Kim:2013mfa,Moon:2012vc}
\begin{eqnarray} \label{SGA}
&&S=S_{\rm E}+S_{\rm TCS} \\
&&S_{\rm E}=\int d^4x
\sqrt{-g}\Big(\frac{R}{2\kappa}-\Lambda\Big),\\
\label{TCS}&&S_{\rm TCS}=-\int d^4x\sqrt{-g}\Big[
\partial_\mu\phi_1\partial^\mu\phi_3+\frac{1}{2}\partial_\mu\phi_2\partial^\mu\phi_2
                  +\mu^2\phi_1\phi_2
                  +m^2\phi_1\phi_3+\frac{1}{2}m^2\phi^2_2\Big],
\end{eqnarray}
where $S_{\rm E}$ is introduced to provide de Sitter background with
$\Lambda>0$ and $S_{\rm TCS}$ represents the three-coupled scalar
theory. Here we have $\kappa=8\pi G=1/M^2_{\rm P}$, $M_{\rm P}$
being the reduced Planck mass, $m^2$ is the degenerate mass-squared,
and $\mu^2$ is a parameter. We follow the conventions
in~\cite{Baumann:2009ds} to compute the power spectrum.

 The Einstein
equation is given by
\begin{equation} \label{ein-eq}
G_{\mu\nu} +\kappa \Lambda g_{\mu\nu}=2\kappa T_{\mu\nu}
\end{equation}
with the energy-momentum tensor
\begin{eqnarray}
T_{\mu\nu}&=&\partial_{\mu}\phi_1\partial_{\nu}\phi_3
+\frac{1}{2}\partial_{\mu}\phi_2\partial_{\nu}\phi_2
\nonumber\\
&&\hspace*{2em}-\frac{1}{2}g_{\mu\nu}\Big(
\partial_{\rho}\phi_1\partial^{\rho}\phi_3
+\frac{1}{2}(\partial_{\rho}\phi_2)^2+\mu^2\phi_1\phi_2+m^2\phi_1\phi_3
+\frac{1}{2}m^2\phi_2^2\Big).\nonumber
\end{eqnarray}
Three scalar equations are obtained  when one varies the action
(\ref{TCS}) with respect to $\phi_3,\phi_2,\phi_1$, respectively,
\begin{equation} \label{b-eq1}
(\nabla^2-m^2)\phi_1=0,~~(\nabla^2-m^2)\phi_2=\mu^2\phi_1,~~(\nabla^2-m^2)\phi_3=\mu^2\phi_2
\end{equation}
which are arranged to give  degenerate fourth-order  and sixth-order
equations
\begin{equation} \label{b-eq2}
(\nabla^2-m^2)^2\phi_2=0,~~(\nabla^2-m^2)^3\phi_3=0.
\end{equation}
It shows how   a higher-derivative scalar theory comes out from the
second-order coupled action (\ref{TCS}). This is so because of the
presence of  $\mu^2\phi_1\phi_2$-term in (\ref{TCS}). We have always
the same second-order equation $(\nabla^2-m^2)\phi_p=0$ for
$p=1,2,3$ without it.

When one chooses the
 vanishing scalars, the  dS spacetime solution is given by
 \begin{equation}
\bar{\phi}_1=\bar{\phi}_2=\bar{\phi}_3=0\to  \bar{R}=4\kappa
\Lambda.
 \end{equation}
Curvature quantities are given by
\begin{equation}
\bar{R}_{\mu\nu\rho\sigma}=H^2(\bar{g}_{\mu\rho}\bar{g}_{\nu\sigma}-\bar{g}_{\mu\sigma}\bar{g}_{\nu\rho}),~~\bar{R}_{\mu\nu}=3H^2\bar{g}_{\mu\nu}
\end{equation}
with a constant Hubble parameter $H^2=\kappa \Lambda/3$. We
represent the dS spacetime explicitly by choosing a conformal time
$\eta$ as a flat slicing
\begin{eqnarray} \label{frw}
ds^2_{\rm dS}=\bar{g}_{\mu\nu}dx^\mu
dx^\nu=a(\eta)^2\Big[-d\eta^2+d{\bf x}\cdot d{\bf x}\Big]
\end{eqnarray}
with  the conformal  scale factor
\begin{eqnarray}
a(\eta)=-\frac{1}{H\eta}\to a(t)=e^{Ht},
\end{eqnarray}
where the latter represents  the scale factor for  cosmic time $t$.
During the dS stage, the scale factor $a$ goes from small to a very
large value like $a_f/a_i\simeq 10^{30}$. It implies that the
conformal time $\eta=-(aH)^{-1}$ runs from $-\infty$ (infinite past)
to $0^-$ (infinite future). As is shown in Fig. 1, the UV/IR
boundaries (${\rm
\partial dS}_{\infty/0}$) of dS space are located   at
$\eta=-\infty$ and  $\eta=0^-$, respectively,  which make the
boundary compact~\cite{Maldacena:2002vr}. Also we recall that this
coordinate system covers only half of dS space and thus,
$\eta=-\infty$  corresponds to the past horizon. We emphasize that
the BD vacuum must be chosen at $\eta=-\infty$, while the dual LCFT
can be thought of as living on the  slice ($\mathbb{R}^3$) at
$\eta=-\epsilon(0<\epsilon \ll1)$. This indicates that one has  to
take into account both boundaries of $\eta=-\infty$ and $-\epsilon$
to compute the power spectrum. This might imply the dS/QFT and
dS/LCFT correspondences.
\begin{figure*}[t!]
\centering
\includegraphics[width=.6\linewidth,origin=tl]{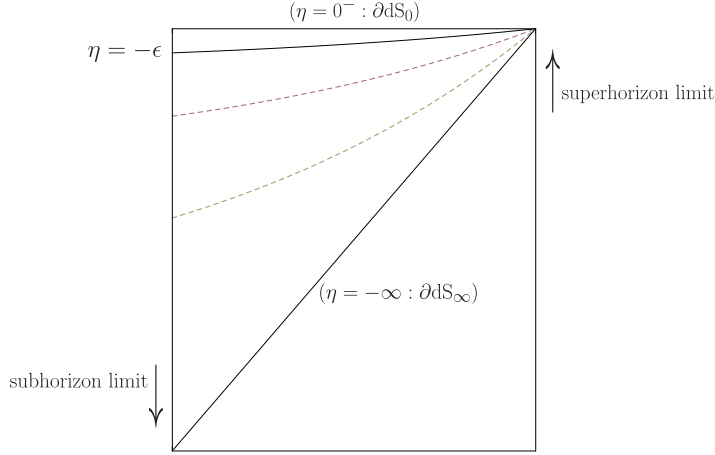}
\caption{Penrose diagram of de Sitter inflation with the UV/IR
boundaries (${\rm
\partial dS}_{\infty/0}$) located   at
$\eta=-\infty$ and  $\eta=0^-$. A slice ($\mathbb{R}^3$) at
$\eta=-\epsilon$ is employed to define the LCFT.  Conformal
invariance in $\mathbb{R}^3$ at $\eta=-\epsilon$ is connected to the
isometry group SO(1,4) of dS space. The dS isometry group acts as
conformal group when fluctuations are superhorizon. Hence,
correlators are expected to be constrained  by conformal invariance.
On the other hand, one introduces the BD vacuum in the subhorizon
limit of $\eta\to -\infty$ to compute the power spectra. }
\end{figure*}

For simplicity, we take   the Newtonian gauge of $B=E=0 $ and
$\bar{E}_i=0$ for metric perturbation $h_{\mu\nu}$ around the dS
background $\bar{g}_{\mu\nu}$ (\ref{frw}). Then, the perturbed
metric is given by
\begin{eqnarray} \label{so3-met}
ds^2=a(\eta)^2\Big[-(1+2\Psi)d\eta^2+2\Psi_i d\eta
dx^{i}+\Big\{(1+2\Phi)\delta_{ij}+h_{ij}\Big\}dx^idx^j\Big]
\end{eqnarray}
with transverse-traceless tensor $\partial_ih^{ij}=h=0$. Also, one
has three scalar perturbations \begin{equation} \phi_p=
\bar{\phi}_p+\varphi_p,~~p=1,2,3.
\end{equation}
In order to obtain  the cosmological perturbed equations, one has to
linearize the Einstein equation (\ref{ein-eq})  directly around the
dS spacetime, arriving at
\begin{eqnarray}
\delta R_{\mu\nu}(h)-3H^2h_{\mu\nu}=0 \to
\bar{\nabla}^2h_{ij}=0.\label{heq}
\end{eqnarray}
Two scalars $\Psi$ and $\Phi$ are not physically propagating modes.
$\Psi=-\Phi$ was found~\cite{Baumann:2009ds} when using the
linearized Einstein equation, and  it was   used to define the
comoving curvature perturbation in the slow-roll inflation.   Also,
a vector $\Psi_i$ is nonpropagating mode since it has no kinetic
term. In the dS inflation, there is no coupling between
$\{\Psi,\Phi\}$ and $\{\varphi_p\}$ because of vanishing background
$\bar{\phi}_p=0$.
  The linearized scalar
equations are given by
\begin{eqnarray}
&&\label{sing-eq1}(\bar{\nabla}^2-m^2)\varphi_1=0,\\
&&\label{sing-eq2}(\bar{\nabla}^2-m^2)\varphi_2=\mu^2\varphi_1, \\
&&\label{sing-eq3}(\bar{\nabla}^2-m^2)\varphi_3=\mu^2\varphi_2.
\end{eqnarray}
These are combined to give  a degenerate fourth-order equation and
and a sixth-order  equation
\begin{eqnarray} \label{tct-eq1}
&&(\bar{\nabla}^2-m^2)^2\varphi_2=0,\\
\label{tct-eq2}&&(\bar{\nabla}^2-m^2)^3\varphi_3=0,
\end{eqnarray}
which are  our main equations to be solved for cosmological purpose
because a complete solution to a second-order equation
(\ref{sing-eq1}) was given by the Hankel function.

\section{dS/LCFT correspondence}
Conformal invariance on the slice ($\mathbb{R}^3$) near $\eta=0^-$
is connected to the isometry group SO(1,4) of dS spacetime. The dS
isometry group acts as conformal group when fluctuations are in the
superhorizon limit of $\eta\to0^-$. The two-point functions
(correlators) are expected to be constrained by conformal
invariance. For definiteness, we first consider the slice
($\mathbb{R}^3$) and its momentum space at $\eta=-\epsilon$ and
then, take the limit of $\epsilon\to 0$~\cite{Kehagias:2012pd}.

To derive  the dS/LCFT correspondence, we first solve Eqs.
(\ref{sing-eq1}), (\ref{tct-eq1}), and (\ref{tct-eq2}) in the
superhorizon limit of $\eta\to 0^-$. Their solutions are given by
\begin{equation} \label{su-asym}
\varphi_{1,\eta\to 0}(\eta) \sim
(-\eta)^w,~~\varphi_{2,\eta\to0}(\eta) \sim
(-\eta)^w\ln[-\eta],~~\varphi_{3,\eta\to0}(\eta) \sim
(-\eta)^w\ln^2[-\eta]
\end{equation}
with
\begin{equation}
w=\frac{3}{2}\Bigg(1-\sqrt{1-\frac{4m^2}{9H^2}}\Bigg).
\end{equation}
In the dS/CFT picture, the complementary series of
$0<\frac{4m^2}{9H^2}<1$ have a dual interpretation in terms of a
unitary CFT while the principal series of $\frac{4m^2}{9H^2}>1$
require a nonunitary CFT~\cite{Strominger:2001pn}. Hence, we choose
the complementary series for developing the dS/LCFT correspondence.
These solutions all satisfy the Dirichlet boundary condition of
$\lim_{\eta \to 0^-}[\varphi_{p,0}] \to 0$.

Deriving cosmological correlator of a  massive scalar
from the dS/CFT dictionary, it is very important to note the following two  statements~\cite{Harlow:2011ke}:\\
 (i) In Lorentzian dS$_{4}$, the extrapolated bulk
 correlators  are  a sum of two contributions. One is the leading
 behavior of a CFT correlator of an operator with conformal dimension
 $w=\frac{3}{2}-\sqrt{\frac{9}{4}-\frac{m^2}{H^2}}$, while the other
comes from the leading
 behavior of a CFT correlator of an operator with dimension
 $\triangle_+=\frac{3}{2}+\sqrt{\frac{9}{4}-\frac{m^2}{H^2}}$. \\
(ii) In Lorentzian dS$_{4}$, functional derivatives of late-time
Schr\"{o}dinger wavefunction produce CFT correlators with dimension
$\triangle_+$ only. \\
The dominant term in (i) was computed by Witten for a particular
scalar~\cite{Witten:2001kn},  whereas  a massless version of
statement (ii) was firstly made by
Maldacena~\cite{Maldacena:2002vr}.  This indicates that the dS/CFT
``extrapolate" and ``differentiate" dictionaries are inequivalent to
each other, while the AdS/CFT ``extrapolate" and ``differentiate"
dictionaries are equivalent.  Following  (ii) to compute
cosmological correlator of a massive scalar, it is inversely
proportional to  CFT correlator with dimension $\triangle_+$ as
\begin{equation}
 \langle\phi(k)\phi(-k)\rangle \propto \frac{1}{\langle {\cal O}(k){\cal
 O}(-k)\rangle}\propto \frac{1}{k^{-3+2\triangle_+}}=k^{2w-3},\end{equation}
which leads to the power spectrum for a massive scalar in the
superhorizon limit.   On the other hand, the cosmological correlator
is directly proportional to the CFT correlator with different
dimension $w$ when one follows (i)
\begin{equation} \label{dir-rel}
\langle\phi(k)\phi(-k)\rangle \propto~ _{\rm
e}\langle\sigma(k)\sigma(-k)\rangle_{\rm e} \propto k^{2w-3}.
\end{equation}

If one uses (i) to derive LCFT-correlators, they are derived from
the relation
\begin{equation}\label{erel}
 _{\rm e}\langle {\cal O}_{p}({\bf x}){\cal
O}_{q}({\bf y})\rangle_{\rm e}=\lim_{\eta,\eta'\to
0}[\eta\eta']^{-w}\langle \varphi_p({\bf x},\eta) \varphi_q({\bf
y},\eta')\rangle,
\end{equation}
where $\langle \varphi_p({\bf x},\eta) \varphi_q({\bf
y},\eta')\rangle$ are Green's functions and its derivative with
respect to $w$. We have derived them in Appendix A explicitly.

 Following (ii) to derive LCFT correlators, one must use the
bulk-to-boundary propagators and relation
\begin{equation}\label{extra-c} \langle {\cal O}_{\bar{p}}({\bf x}){\cal
O}_{\bar{q}}({\bf y})\rangle=-\frac{\delta^2 \ln Z_{\rm
bulk}}{\delta\varphi_{p,0}({\bf x})\delta\varphi_{q,0}({\bf y})},
\end{equation}
where $\bar{3}=1,\bar{2}=2,\bar{1}=3$ and
\begin{equation}
Z_{\rm bulk}=e^{-\delta S_{\rm Sb}[\{\varphi_{a,0}\}]}.
\end{equation}
See Appendix B for detail derivations using ``differentiate"
dictionary. Explicitly, rank-3 LCFT correlators are determined by
\begin{eqnarray}
&& \label{cc0} \langle {\cal O}_1({\bf x}){\cal O}_1({\bf
y})\rangle=~\langle {\cal O}_1({\bf x}){\cal O}_2({\bf
y})\rangle=0,\\
\label{cc1} &&\langle {\cal O}_1({\bf x}){\cal O}_3({\bf
y})\rangle=~\langle {\cal O}_2({\bf x}){\cal O}_2({\bf
y})\rangle=\frac{A}{|{\bf x}-{\bf
y}|^{2\triangle_+}}, \\
&&  \label{cc2}\langle {\cal O}_2({\bf x}){\cal O}_3({\bf
y})\rangle=\frac{A}{|{\bf x}-{\bf y}|^{2\triangle_+}}\Big(-2\ln|{\bf
x}-{\bf y}|+D_1\Big),\\
 &&  \label{cc3}\langle {\cal O}_3({\bf
x}){\cal O}_3({\bf y})\rangle=\frac{A}{|{\bf x}-{\bf
y}|^{2\triangle_+}}\Big(2\ln^2|{\bf x}-{\bf y}|-2D_1\ln|{\bf x}-{\bf
y}|+D_2\Big) ,
\end{eqnarray}
where constants $A$, $D_{1}$, and  $D_2$ are given by
\begin{equation}
A=c_0\triangle_+,~~D_1=\frac{1}{\triangle_+}+\frac{1}{c_0}\frac{\partial
c_0}{\partial
\triangle_+},~~D_2=\frac{1}{c_0\triangle_+}\frac{\partial
c_0}{\partial \triangle_+} +\frac{1}{2c_0}\frac{\partial^2
c_0}{\partial \triangle_+^2}.
\end{equation}
We note here that $c_0$ is an undetermined constant. 
Eqs.(\ref{cc0})-(\ref{cc3}) are summarized to be
schematically~\cite{Kim:2013waf,Kim:2013mfa}
\begin{equation} \label{lcft-mat}
\langle {\cal O}_p({\bf x}){\cal O}_q({\bf y})\rangle =\left(
   \begin{array}{ccc}
     0 &0& {\rm CFT} \\
     0&{\rm CFT} & {\rm L} \\
     {\rm CFT} &{\rm L} & {\rm L^2} \\
   \end{array}
 \right),
 \end{equation}
where CFT, L , and  L$^2$  represent their correlators in
(\ref{cc1}), (\ref{cc2}), and (\ref{cc3}), respectively.

 In order to
derive  LCFT correlators in momentum space, one may  use the
relation
\begin{equation}\label{rela0}
\frac{1}{|{\bf x}-{\bf
y}|^{2\triangle_+}}=\frac{\Gamma(\frac{3}{2}-\triangle_+)}{4^{\triangle_+}
\pi^{3/2}\Gamma(\triangle_+)}\int d^3{\bf k}|{\bf
k}|^{2\triangle_+-3}e^{i{\bf k}\cdot ({\bf x}-{\bf
y})},\end{equation} where we observe an inverse-relation of exponent
$2\triangle_+$ between $|{\bf x}|$ and $k=|{\bf k}|$. However, it
seems difficult to derive  momentum correlators of (\ref{cc2}) and
(\ref{cc3}) because of the presence of log-terms. Instead,
following~\cite{Kehagias:2012pd}, we obtain them newly
\begin{eqnarray}
 \label{m0} \langle {\cal O}_1({\bf k}_1){\cal O}_1({\bf
k}_2)\rangle'&=& \langle {\cal O}_1({\bf k}_1){\cal O}_2({\bf
k}_2)\rangle'=0,\\
\label{m1} \langle {\cal O}_1({\bf k}_1){\cal O}_3({\bf
k}_2)\rangle'&=&\langle {\cal O}_2({\bf k}_1){\cal O}_2({\bf
k}_2)\rangle'=\frac{A_0}{k_1^{3-2\triangle_+}}, \\
 \label{m2}\langle {\cal O}_2({\bf k}_1){\cal O}_3({\bf k}_2)\rangle'
&=&a\langle {\cal O}_1({\bf k}_1){\cal O}_3({\bf
k}_2)\rangle'+\frac{\partial}{\partial \triangle_+}\langle {\cal
O}_1({\bf k}_1){\cal O}_3({\bf k}_2)\rangle'\nonumber
\\
&=&\frac{A_0}{k_1^{3-2\triangle_+}}\Big[2\ln[k_1]+a+\frac{A_{0,\triangle_+}}{A_0}\Big],\\
\label{m3}\langle {\cal O}_3({\bf k}_1){\cal O}_3({\bf k}_2)\rangle'
&=&a\langle {\cal O}_1({\bf k}_1){\cal O}_3({\bf
k}_2)\rangle'+b\frac{\partial}{\partial \triangle_+}\langle {\cal
O}_1({\bf k}_1){\cal O}_3({\bf
k}_2)\rangle'+\frac{1}{2}\frac{\partial^2}{\partial
\triangle_+^2}\langle {\cal O}_1({\bf k}_1){\cal O}_3({\bf
k}_2)\rangle'\nonumber
\\
&=&\frac{A_0}{k_1^{3-2\triangle_+}}\Big[2\ln^2[k_1]
+2\Big(b+\frac{A_{0,\triangle_+}}{A_0}\Big)\ln[k_1]
+a+b\frac{A_{0,\triangle_+}}{A_0}+\frac{1}{2}\frac{A_{0,\triangle_+\triangle_+}}{A_0}\Big],\nonumber\\
\end{eqnarray}
where the prime ($'$) denotes the correlators without the
$(2\pi)^3\delta^3(\Sigma_i{\bf k}_i)$ and $a,~b$ are arbitrary
constants. Also,  $A_0$ is given by
\begin{equation}
A_0=\frac{A\Gamma(\frac{3}{2}-\triangle_+)}{4^{\triangle_+}
\pi^{3/2}\Gamma(\triangle_+)}=\frac{c_0\triangle_+\Gamma(\frac{3}{2}-\triangle_+)}{4^{\triangle_+}
\pi^{3/2}\Gamma(\triangle_+)}
\end{equation}
which was obtained from using the relation (\ref{rela0}) together
with (\ref{cc1}). Here, $A_{0,\triangle_+}$ and
$A_{0,\triangle_+\triangle_+}$ denote derivatives of $ A_0$ and
$A_{0,\triangle_+}$ with respect to $\triangle_+$, respectively.
These correlators will be compared to the power spectra obtained  in
the superhorizon limit of $z\to 0$ in Sec. 5 by choosing $a$ and $b$
appropriately. Actually, there is  ambiguity for fixing $D_1$ and
$D_2$ in (\ref{cc2}) and (\ref{cc3}). It implies that these depend
on the computation scheme. For example, these are given by $\zeta_1$
and $\zeta_2$ in (\ref{g1e}) and (\ref{g2e}) when using the
``extrapolate" dictionary. Including $a$ and $b$ in (\ref{m2}) and
(\ref{m3}) reflects this ambiguity.

 Finally, to compare (\ref{m0})-(\ref{m3})  with the power spectra, we express
 LCFT-correlators as
 \begin{equation} \label{mlcft-c}
\langle {\cal O}_p({\bf k}_1){\cal O}_q({\bf k}_2)\rangle'=\langle
{\cal O}_2({\bf k}_1){\cal O}_2({\bf k}_2)\rangle'\times\langle
{\cal O}_p({\bf k}_1){\cal O}_q({\bf k}_2)\rangle'_{\rm L},
\end{equation}
where $\langle {\cal O}_p({\bf k}_1){\cal O}_q({\bf
k}_2)\rangle'_{\rm L}$ are contributions from logarithmic parts.

\section{Three scalar propagations in dS}

In order to calculate  the power spectrum, we have to know the
solution to  Eqs. (\ref{sing-eq1}), (\ref{tct-eq1}), and
(\ref{tct-eq2}) in the whole range of $\eta(z)$. Also,  these
solutions are required to
 satisfy two coupled equations (\ref{sing-eq2}) and (\ref{sing-eq3})
simultaneously. For cosmological  purpose, the scalars $\varphi_{p}$
can be expanded in Fourier modes $\phi^{p}_{\bf k}(\eta)$
\begin{eqnarray}\label{scafou}
\varphi_{p}(\eta,{\bf x})=\frac{1}{(2\pi)^{\frac{3}{2}}}\int d^3{\bf
k}~\phi^{p}_{\bf k}(\eta)e^{i{\bf k}\cdot{\bf x}}.
\end{eqnarray}
The second-order  equation (\ref{sing-eq1}) leads to
\begin{eqnarray}\label{scalar-eq2}
\Bigg[\frac{d^2}{d \eta^2}-\frac{2}{\eta}\frac{d}{d
\eta}+k^2+\frac{m^2}{H^2}\frac{1}{\eta^2}\Bigg]\phi^1_{\bf
k}(\eta)=0
\end{eqnarray}
which could be expressed  in term of  $z=-k\eta$
\begin{equation}\label{scalar-eq3}
\Big[\frac{d^2}{dz^2}-\frac{2}{z}\frac{d}{dz}+1+\frac{m^2}{H^2}\frac{1}{z^2}\Big]\phi^{1}_{\bf
k}(z)=0.
\end{equation}
 The solution to (\ref{scalar-eq3}) is given by
the Hankel function $H^{(1)}_\nu$ as
\begin{equation} \label{scalar-eq5}
\phi^1_{\bf k}(z)=\frac{H}{\sqrt{2k^3}}
\sqrt{\frac{\pi}{2}}e^{i(\frac{\pi\nu}{2}+\frac{\pi}{4})}z^{3/2}H^{(1)}_{\nu}(z),~~\nu=\sqrt{\frac{9}{4}-\frac{m^2}{H^2}}.
\end{equation}
In the subhorizon limit of $z\to \infty$, Eq.(\ref{scalar-eq3})
reduces to
\begin{equation}\label{scalar-eq6}
\Big[\frac{d^2}{dz^2}-\frac{2}{z}\frac{d}{dz}+1\Big]\phi^{1}_{{\bf
k},\infty}(z)=0
\end{equation}
which implies  the positive-frequency  solution with the
normalization $1/\sqrt{2k}$
\begin{equation} \label{scalar-eq7}
\phi^{1}_{{\bf k},\infty}(z)=\frac{H}{\sqrt{2k^3}}(i+z)e^{iz}.
\end{equation}
This is also  a typical mode solution of a massless scalar
propagating on whole dS spacetime. In the superhorizon limit of
$z\to0$, Eq.(\ref{scalar-eq3}) takes the form
\begin{equation}\label{scalar-eq11}
\Bigg[\frac{d^2}{dz^2}-\frac{2}{z}\frac{d}{dz}+\frac{m^2}{H^2}\frac{1}{z^2}\Bigg]\phi^1_{{\bf
k},0}(z)=0,
\end{equation}
whose solution is given by
\begin{equation}
\phi^1_{{\bf
k},0}(z)=\frac{H}{\sqrt{2k^3}}z^{w},~~w=\frac{3}{2}-\nu.
\end{equation}

On the other hand,  plugging (\ref{scafou}) into (\ref{tct-eq1})
leads to a degenerate fourth-order  equation for $\phi^2_{\bf
k}(\eta)$
\begin{eqnarray}
\Bigg[\eta^2\frac{d^2}{d\eta^2}-2\eta\frac{d}{d\eta}+k^2\eta^2+\frac{m^2}{H^2}\Bigg]^2\phi^2_{\bf
k}(\eta)=0\label{ss2-eq2}
\end{eqnarray}
which seems difficult to be solved directly. However, we may solve
Eq.(\ref{ss2-eq2}) in  the two limits of subhorizon and
superhorizon. In the subhorizon limit of $z\to \infty$,
Eq.(\ref{ss2-eq2}) takes the form
\begin{equation}\label{sub-eq1}
\Bigg[z^2\frac{d^2}{dz^2}-2z\frac{d}{dz}+z^2\Bigg]^2\phi^2_{{\bf
k},\infty}=0.
\end{equation}
whose  direct solution is given by
\begin{eqnarray} \label{sub-sol}
\phi^{2,d}_{{\bf k},\infty}=\tilde{c}_2(i+z)e^{iz}.
\end{eqnarray}
 The complex conjugate  of $\phi^{2,d}_{{\bf k},\infty}$ is a solution to
(\ref{sub-eq1}) too.  Importantly, we note that Eq.(\ref{sing-eq2})
reduces to a second-order equation  in the subhorizon limit
\begin{equation}
\Big[\frac{d^2}{dz^2}-\frac{2}{z}\frac{d}{dz}+1\Big]\phi^{2}_{{\bf
k},\infty}(z)=0,
\end{equation}
whose solution is also given by \begin{equation} \label{phi2s-sol}
\phi^{2}_{{\bf k},\infty}(z)=\tilde{c}_2(i+z)e^{iz}=\phi^{2,d}_{{\bf
k},\infty}(z).
\end{equation}

Curiously, Eq.(\ref{ss2-eq2}) takes the form in the superhorizon
limit of $z\to 0$ as
\begin{eqnarray}
\Bigg[z^2\frac{d^2}{dz^2}-2z\frac{d}{dz}+\frac{m^2}{H^2}\Bigg]^2\phi^2_{{\bf
k},0}(z)=0.\label{super-eq2}
\end{eqnarray}
Its solution is given by the log-function
\begin{equation}\label{super-phi2}
\phi^2_{{\bf k},0}(z)= z^w\ln[z].
\end{equation}
The presence of ``$\ln[z]$" reflects  that (\ref{super-phi2}) is a
solution to the fourth-order equation (\ref{super-eq2}). For
$\phi^1_{{\bf k},0}(z)= z^w$, $\phi^2_{{\bf k},0}(z)$ also satisfies
the superhorizon limit of a coupled equation (\ref{sing-eq2})
\begin{eqnarray}
\Bigg[z^2\frac{d^2}{dz^2}-2z\frac{d}{dz}+\frac{m^2}{H^2}\Bigg]\phi^2_{{\bf
k},0}(z)=-\frac{\mu^2}{H^2}\phi^1_{{\bf k},0}(z)\label{supersub-eq2}
\end{eqnarray}
 for having a choice of $\mu^2=(3-2w)H^2$.

Lastly, we have the degenerate sixth-order equation for $\phi^3_{\bf
k}(\eta)$
\begin{eqnarray}
\Big[\eta^2\frac{d^2}{d\eta^2}-2\eta\frac{d}{d\eta}+k^2\eta^2+\frac{m^2}{H^2}\Big]^3\phi^3_{\bf
k}(\eta)=0\label{s2-eq2}
\end{eqnarray}
which seems formidable to be solved exactly. However, its equations
in the subhorizon  limit takes the form
\begin{eqnarray}
\Bigg[z^2\frac{d^2}{dz^2}-2z\frac{d}{dz}+z^2\Bigg]^3\phi^3_{{\bf
k},0}(z)=0.\label{sub-eq3}
\end{eqnarray}
A direct solution is given by
\begin{equation}\label{sub-phi3}
\phi^{3,d}_{{\bf k},0}(z)=\tilde{c}_3(i+z)e^{iz}.
\end{equation}
Eq.(\ref{sing-eq3}) reduces to a second-order equation  in the
subhorizon limit
\begin{equation}
\Big[\frac{d^2}{dz^2}-\frac{2}{z}\frac{d}{dz}+1\Big]\phi^{3}_{{\bf
k},\infty}(z)=0,
\end{equation}
whose solution is  given by \begin{equation} \label{phi3s-sol}
\phi^{3}_{{\bf k},\infty}(z)=\tilde{c}_3(i+z)e^{iz}=\phi^{3,d}_{{\bf
k},\infty}(z).
\end{equation}

In the superhorizon limit, Eq.(\ref{s2-eq2}) leads to
\begin{eqnarray}
\Bigg[z^2\frac{d^2}{dz^2}-2z\frac{d}{dz}+\frac{m^2}{H^2}\Bigg]^3\phi^3_{{\bf
k},0}(z)=0\label{super-eq3}
\end{eqnarray}
whose solution is given by (see Appendix C for derivation using the
trick in~\cite{Kogan:1999bn})
\begin{equation}\label{super-phi3}
\phi^3_{{\bf k},0}(z)\propto z^w\ln^2[z].
\end{equation}
 Here, the presence of ``$\ln^2[z]$" indicates
that (\ref{super-phi3}) is a solution to the degenerate sixth-order
equation (\ref{super-eq3}).  Explicitly, one has three steps to show
that $\phi^3_{{\bf k},0}(z)$ is a solution to (\ref{super-eq3})
\begin{eqnarray}
\label{super-eq41}&&\Big[z^2\frac{d^2}{dz^2}-2z\frac{d}{dz}+\frac{m^2}{H^2}\Big]\phi^3_{{\bf
k},0}(z)\to 2(2w-3)z^w\ln[z]+2z^w,\\
\label{super-eq42}&&\Big[z^2\frac{d^2}{dz^2}-2z\frac{d}{dz}+\frac{m^2}{H^2}\Big]^2\phi^3_{{\bf
k},0}(z)\to 2(2w-3)^2z^w,\\
\label{super-eq43}&&\Big[z^2\frac{d^2}{dz^2}-2z\frac{d}{dz}+\frac{m^2}{H^2}\Big]^3\phi^3_{{\bf
k},0}(z)\to 0.
\end{eqnarray}
We point out  that considering $\phi^2_{{\bf k},0}(z)= z^w\ln[z]$,
$\phi^3_{{\bf
k},0}(z)=\frac{z^w\ln^2[z]}{2}-\frac{1}{2w-3}z^w\ln[z]$ also
satisfies the superhorizon limit of a coupled equation
(\ref{sing-eq3})
\begin{eqnarray}
\Bigg[z^2\frac{d^2}{dz^2}-2z\frac{d}{dz}+\frac{m^2}{H^2}\Bigg]\phi^3_{{\bf
k},0}(z)=-\frac{\mu^2}{H^2}\phi^2_{{\bf k},0}(z)\label{supersub-eq3}
\end{eqnarray}
 by choosing  $\mu^2=(3-2w)H^2$.

Consequently, we summarize the two asymptotic solutions. The
solutions are given by the same form in the subhorizon limit,
irrespective of their higher-order derivative equations, as
\begin{equation} \label{z-infinity}
\phi^{p}_{{\bf k},\infty}(z)=\tilde{c}_p(i+z)e^{iz}=\phi^{p,d}_{{\bf
k},\infty}(z),
\end{equation}
while these take different forms in the superhorizon limit
\begin{equation}\label{solsu}
\phi^{1}_{{\bf k},0}(z)=z^w,~~\phi^{2}_{{\bf
k},0}(z)=z^w\ln[z],~~\phi^3_{{\bf
k},0}(z)=\frac{z^w\ln^2[z]}{2}-\frac{1}{2w-3}z^w\ln[z].
\end{equation}
This implies that the solution feature to the higher-order
derivative equation appears in the superhorizon region only, but the
solution  to the second-order equation always appears in the
subhorizon region. This is because we are not interested in
(\ref{tct-eq1}) and (\ref{tct-eq2}), but rather in (\ref{sing-eq2})
and (\ref{sing-eq3}) where the right-handed side is subdominant in
the subhorizon limit.  The former solution will be  used to define
the dual LCFT via the dS/LCFT correspondence, while the latter will
be exploited to define the BD vacuum for quantum fluctuations
through the dS/QFT correspondence.

\section{Power spectra}

The power spectrum is  defined by the two-point  function.  The
defining relation is given by
\begin{equation}
_{\rm
BD}\langle0|\varphi_p(\eta,\bold{x})\varphi_q(\eta,\bold{y})|0\rangle_{\rm
BD}=\int d^3{\bf k}\Big[ \frac{{\cal P}_{pq}(k,\eta)}{4\pi
k^3}\Big]e^{i \bold{k}\cdot (\bold{x}-\bold{y})},
\end{equation}
where $k=|{\bf k}|$ is the comoving wave number. It could be
computed when one chooses   the BD vacuum state $|0\rangle_{\rm BD}$
which is the Minkowski vacuum of a comoving observer in the distant
past [in the subhorizon limit of
 $\eta\to -\infty(z\to \infty)]$ when the
mode is deep inside the horizon~\cite{Baumann:2009ds}. Quantum
fluctuations were created on all length scales with wave number $k$.
Cosmologically relevant fluctuations start their lives deep inside
the comoving Hubble radius $(aH)^{-1}$ which defines the subhorizon:
$k~\gg aH(z\gg1)$. On later, the comoving Hubble radius shrinks
during inflation while keeping the wavenumber $k$ constant. All
fluctuations exit the comoving Hubble radius, they reside on  the
superhorizon region of $k~\ll aH(z\ll1)$ after horizon crossing.
 In the dS inflation, we choose the subhorizon limit
of  $z\to \infty$ (the UV boundary)  to define the BD vacuum, while
the superhorizon limit (the IR boundary) is chosen as $z\to 0$ to
define the dS/LCFT correspondence.

 To compute the power spectrum, we have to know the commutation relations and the Wronskian conditions. The canonical
conjugate momenta are given by
\begin{equation}
\pi_1=a^2\frac{d\varphi_3}{d\eta},~~\pi_2=a^2\frac{d\varphi_2}{d\eta},~~\pi_3=a^2\frac{d\varphi_1}{d\eta},
\end{equation}
where the mid-term is considered as a standard canonical momentum.
The canonical quantization is accomplished by imposing equal-time
commutation relations:
\begin{eqnarray}\label{comm}
[\hat{\varphi}_{p}(\eta,{\bf x}),\hat{\pi}_{q}(\eta,{\bf
y})]=i\delta_{pq}\delta^3({\bf x}-{\bf y}).
\end{eqnarray}
 The three operators $\hat{\varphi}_{p}$ are expanded in terms of Fourier modes
as~\cite{Kim:2013waf,Kim:2013mfa}
\begin{eqnarray}\label{hex1}
\hat{\varphi}_{1}(z,{\bf x})&=&\frac{1}{(2\pi)^{\frac{3}{2}}}\int
d^3{\bf k}N_1\Big[\Big(i\hat{a}_1({\bf k})\phi^1_{\bf k}(z)e^{i{\bf
k}\cdot{\bf
x}}\Big)+{\rm h.c.}\Big], \\
\label{hex2} \hat{\varphi}_2(z,{\bf
x})&=&\frac{1}{(2\pi)^{\frac{3}{2}}}\int d^3{\bf
k}N_2\Big[\Big(\hat{a}_2({\bf k})\phi^1_{\bf k}(z)+\hat{a}_1({\bf
k})\phi^2_{\bf k}(z)\Big)e^{i{\bf k}\cdot{\bf
x}}+{\rm h.c.}\Big],\\
\label{hex3}\hat{\varphi}_3(z,{\bf
x})&=&\frac{1}{(2\pi)^{\frac{3}{2}}}\int d^3{\bf k}N_3
          \Big[i\Big\{\hat{a}_3({\bf k})\phi^1_{\bf
k}(z)+\Big(\hat{a}_2({\bf k})-\frac{i}{2}\hat{a}_1({\bf
k})\Big)\phi^2_{\bf k}(z)\\ \nonumber
                      &+&\frac{1}{2}\hat{a}_1({\bf k})\phi^3_{\bf
k}(z)\Big\}e^{i{\bf k}\cdot{\bf x}}+{\rm h.c.}\Big].
\end{eqnarray}
with $\{N_{p}\}$ the normalization constants.  Here it is worth
noting that we do not know the complete solutions $\{\phi^p_{\bf
k}(z)\}$ because we could not solve the degenerate fourth-order
equation (\ref{ss2-eq2}) and sixth-order equation (\ref{s2-eq2})
completely. However, if one uses the asymptotic solutions
$\phi^{p,0}_{\bf k}=\tilde{c}_p(i+z)e^{iz}$ in the subhorizon limit
instead of $\phi^{p}_{\bf k}$, one may impose (\ref{comm}) to derive
the commutation relation between annihilation and creation
operators. Plugging (\ref{hex1})-(\ref{hex3}) into (\ref{comm})
determines the relation of normalization constants as $N_1N_3=-1/2k
$ and $N_2=1/\sqrt{2k}$. Also, the commutation relations between
$\hat{a}_p({\bf k})$ and $\hat{a}^{\dagger}_q({\bf k}')$ are
obtained  to be
 \begin{equation} \label{scft}
 [\hat{a}_p({\bf k}), \hat{a}^{\dagger}_q({\bf k}')]= 2k
 \left(
  \begin{array}{ccc}
   0 &0& -1  \\
    0 & 1& i\\
    -1 & -i& \frac{3}{2}\\
  \end{array}
 \right)\delta^3({\bf k}-{\bf k}'),
 \end{equation}
 which indicates a quantum nature of three-coupled scalar theory.
 This shows the dS/QFT correspondence in the subhorizon limit.
 It is also  noted that a factor of $\frac{3}{2}$ in $[\hat{a}_3({\bf k}),\hat{a}^{\dagger}_3({\bf k}')]$ represents
 higher-derivative nature for $\varphi_3$.

 We
 note that the off-diagonal commutation relations $
[\hat{\varphi}_{p}(\eta,{\bf x}),\hat{\pi}_{q}(\eta,{\bf y})]=0$ for
$p\neq q$ gives the following Wronskian conditions  together with
(\ref{scalar-eq7}), $\tilde{c}_2=iH/(2\sqrt{2k^3})$
(\ref{phi2s-sol}), and $\tilde{c}_3=H/(4\sqrt{2k^3})$
(\ref{phi3s-sol}):
\begin{eqnarray}
&&\hspace*{-2em}a^2\Big(\phi^1_{{\bf
k},\infty}\frac{d\phi^{2*}_{{\bf k},\infty}}{dz}-\phi^{2*}_{{\bf
k},\infty}\frac{d\phi^{1}_{{\rm k},\infty}}{dz}+\phi^{1*}_{{\bf
k},\infty}\frac{d\phi^{2}_{{\bf k},\infty}}{dz}-\phi^{2}_{{\bf
k},\infty}\frac{d\phi^{1*}_{{\rm
k},\infty}}{dz}\Big)=-\frac{1}{k},\label{wron1}\\
&&\hspace*{-2em}\Big(\phi^1_{{\bf k},\infty}\frac{d\phi^{3*}_{{\bf
k},\infty}}{dz}-\phi^{3*}_{{\bf k},\infty}\frac{d\phi^{1}_{{\rm
k},\infty}}{dz}+\phi^{3}_{{\bf k},\infty}\frac{d\phi^{1*}_{{\rm
k},\infty}}{dz}-\phi^{1*}_{{\bf k},\infty}\frac{d\phi^{3}_{{\bf
k},\infty}}{dz}\Big) \nonumber \\
&&=2\Big(\phi^2_{{\bf k},\infty}\frac{d\phi^{2*}_{{\bf
k},\infty}}{dz}-\phi^{2*}_{{\bf k},\infty}\frac{d\phi^{2}_{{\rm
k},\infty}}{dz}\Big).\label{wron2}
\end{eqnarray}

Now  we  are position to choose the BD vacuum $|0\rangle_{\rm BD}$
by imposing $\hat{a}_p({\bf k})|0\rangle_{\rm BD}=0$. We should
explain what the BD vacuum is really, since  the three-coupled
scalar theory is quite different from the three free-scalar theory
without $\mu^2\varphi_1\varphi_2$.  We mention briefly how to
quantize the $n$-coupled scalar field theory within the
Becchi-Rouet-Stora-Tyutin (BRST) quantization scheme in Minkowski
space~\cite{Kim:2013mfa}. It has been carried out by introducing the
FP ghost action composed of $n$-FP ghost fields. Extending a BRST
quartet generated by two scalars and FP ghosts to $n$ scalars and FP
ghosts, there remains a physical subspace with positive norm for odd
$n$, while there exists only the vacuum for even $n$. This has shown
the non-triviality of a odd-higher derivative scalar field theory,
which might show a hint to resolve the nonunitarity confronted when
developing a higher-order derivative quantum gravity. Explicitly,
the $n$ = 2 case corresponds to a dipole ghost field for the
singleton. They have formed a quartet to give the zero norm state
when one includes the FP ghost action, leaving the vacuum only.  On
the other hand, the $n$ = 3 case is enough to have a physical
subspace with positive norm state upon requiring the BRST quartet
mechanism. Comparing it with Yang-Mills theory (4.52)
in~\cite{Kugo:1979gm}, we have an apparent correspondence between
two
\begin{equation}
\varphi_1 \leftrightarrow B,~~\varphi_2 \leftrightarrow A_{\rm
T},~~\varphi_3 \leftrightarrow A_{\rm L},
\end{equation}
where $B$ is a conjugate momentum of scalar gauge mode $A_{\rm S}$,
while $A_{\rm T}$ represents the transverse gauge mode with positive
norm and $A_{\rm L}$ denotes the longitudinal gauge mode with
negative norm.  Additionally, we note a difference arising from a
non-zero commutator of $[\hat{a}_2({\bf k}),\hat{a}^\dagger_3({\bf
k}')]=2i k \delta^3({\bf k}-{\bf k}')$ whose dual  plays an
important role in selecting  a physical CFT $_{\rm e}\langle {\cal
O}_2({\bf x}){\cal O}_2({\bf y})\rangle_{\rm e}$ in the rank-3 LCFT.
This implies that the three-coupled scalar theory provides a
physical scalar field $\varphi_2$ even though it couples to
$\varphi_3$ via (\ref{sing-eq3}).  No larger than $n$ = 3-coupled
scalar theory is necessary to construct a unitary scalar theory from
a higher-derivative scalar theory. Here, the subsidiary condition
(the Gupta-Bleuler condition~\cite{Tolley:2001gg}) of
$\varphi_1^+({\bf x})|$phys$\rangle=0$~\cite{AI} either to find a
physical field with positive norm or to eliminate unphysical field
with negative norm is translated into $\hat{a}_1({\bf
k})|$phys$\rangle=0$ which shares a property of the BD vacuum
$|0\rangle_{\rm BD}$ defined by $\hat{a}_1({\bf k})|0\rangle_{\rm
BD}=0$, in addition to $\hat{a}_2({\bf k})|0\rangle_{\rm BD}=0$ and
$\hat{a}_3({\bf k})|0\rangle_{\rm BD}=0$.

The scalar power spectrum for $\varphi_1$ and ${\cal P}_{\rm
12}(={\cal P}_{\rm 21})$ vanish as
\begin{eqnarray}
{\cal P}_{\rm 11}~=~{\cal P}_{\rm 12}~=~{\cal P}_{\rm 21}~=~0
\end{eqnarray}
when one used the unconventional  relations $[\hat{a}_1({\bf k}),
\hat{a}^{\dagger}_1({\bf k}')]=0,~[\hat{a}_1({\bf k}),
\hat{a}^{\dagger}_2({\bf k}')]=0,$ and $[\hat{a}_2({\bf k}),
\hat{a}^{\dagger}_1({\bf k}')]=0$.

 On the other hand,  the power spectrum of $\varphi_{2}$
and ${\cal P}_{\rm 13}(={\cal P}_{\rm 31})$ are given  by the
conventional massive scalar
\begin{eqnarray}
{\cal P}_{\rm 22}~=~{\cal P}_{\rm 13}~=~{\cal P}_{\rm
31}&=&\frac{k^3}{2\pi^2}\Big|\phi_{\bf k}^1\Big|^2\nonumber\\
&=&\frac{H^2}{8\pi}z^3|e^{i(\frac{\pi
\nu}{2}+\frac{\pi}{4})}H_{\nu}^{(1)}(z)|^2.
\end{eqnarray}
The remaining power spectrum ${\cal P}_{\rm 23}(={\cal P}_{\rm 32})$
and ${\cal P}_{\rm 33}$ are given by
\begin{eqnarray}\label{pw23}
{\cal P}_{\rm 23}=\frac{k^3}{2\pi^2}\Big[|\phi_{\bf
k}^1|^2-i(\phi_{\bf k}^1\phi_{\bf k}^{2*}-\phi_{\bf k}^2\phi_{\bf
k}^{1*})\Big]
\end{eqnarray}
and
\begin{eqnarray}
{\cal P}_{\rm 33}=\frac{k^3}{2\pi^2}\Big[\frac{3}{2}|\phi_{\bf
k}^1|^2-\frac{3i}{2}(\phi_{\bf k}^1\phi_{\bf k}^{2*}-\phi_{\bf
k}^2\phi_{\bf k}^{1*})+\Big|\phi_{\bf
k}^2\Big|^2-\frac{1}{2}(\phi_{\bf k}^1\phi_{\bf k}^{3*}+\phi_{\bf
k}^3\phi_{\bf k}^{1*})\Big],
\end{eqnarray}
 where we fixed $N_3=1/\sqrt{2k}$.

 It is important to note that in the superhorizon
limit of $z\to 0$, ${\cal P}_{\rm 23,0}$ and ${\cal P}_{\rm 33,0}$
are given by
\begin{eqnarray}\label{pw023}
{\cal P}_{\rm 23,0}\to \xi^2z^{2w}\Big(2\ln[z]+1\Big)
\end{eqnarray}
and
\begin{eqnarray}\label{pw033}
{\cal P}_{\rm 33,0}\to
\xi^2z^{2w}\Bigg\{2\ln^2[z]+\frac{6w-11}{2w-3}\ln[z]
+\frac{3}{2}\Bigg\},
\end{eqnarray}
 which implies that ${\cal P}_{\rm 23,0}$ and ${\cal P}_{\rm 23,0}$ approach zero
when $z\to 0$. In deriving (\ref{pw023}) and (\ref{pw033}), $\xi$
was chosen to be  a real quantity given by
\begin{eqnarray}
\phi_{{\bf k},0}^1\sim -i\xi z^{w},~~~\phi_{{\bf k},0}^2\sim -\xi
z^{w}\ln[z],~~~\phi_{{\bf k},0}^3\sim 2i\xi
z^w\Big(\frac{\ln^2[z]}{2}-\frac{\ln[z]}{2w-3}\Big).
\end{eqnarray}
 Consequently, we obtain the whole power spectra in the superhorizon limit of $z=-k\eta\to 0$
\begin{eqnarray} \label{ps-matt1}
{\cal P}_{{ab},0}(k,\eta)&=&\xi^2 \left(
   \begin{array}{ccc}
     0 & 0 & z^{2w} \\
     0 & z^{2w} & z^{2w}(2\ln [z]+1)\\
   z^{2w} & z^{2w}(2\ln [z]+1) &z^{2w}\Big\{2\ln^2[z]+\frac{6w-11}{2w-3}\ln[z]
+\frac{3}{2}\Big\}\end{array}
 \right)
 \end{eqnarray}
 with
 \begin{equation}
 \xi^2=\frac{1}{2^{2w}}\Big(\frac{H}{2\pi}\Big)^2\Bigg(\frac{\Gamma(\frac{3}{2}-w)}{\Gamma(\frac{3}{2})}\Bigg)^2.
 \end{equation}
 For $\eta=-\epsilon(0<\epsilon\ll 1)$~\cite{Larsen:2003pf,Seery:2006tq}, Eq.(\ref{ps-matt1}) takes the
 form
\begin{eqnarray}
 \label{ps-matt3}{\cal P}_{{ab},0}(k,-\epsilon)&=&\xi^2 (\epsilon k)^{2w}\left(
   \begin{array}{ccc}
     0 & 0 &  1\\
     0 & 1 & 2\ln [\epsilon k]+1\\
   1 & 2\ln [\epsilon k]+1 &
   2\ln^2[\epsilon k]+\frac{6w-11}{2w-3}\ln[\epsilon k]
+\frac{3}{2}\end{array}
 \right).
 \end{eqnarray}
Now we are in a position to compare the power spectra
(\ref{ps-matt3}) with LCFT correlators (\ref{mlcft-c}). For this
purpose, we wish to choose $A_0, a,b$ as
\begin{equation}
A_0=e^{(1-a)\triangle_+},~a=(2\triangle_+-5\pm s)/(4\triangle_+-6)
,~b=(4\triangle_+-6\pm s)/(4\triangle_+-6)\end{equation} with
$s=\sqrt{44\triangle_+-12\triangle_+^2-35}$. Then, we observe the
relation
\begin{equation} \label{p-cc}
(2\pi)^4 \pi k^{-3}{\cal P}_{{ab},0}(k,-1)=\langle \phi^a_{{\bf
k}}\phi^b_{-{\bf k}}\rangle   \propto \frac{1}{\langle {\cal
O}_2(k){\cal O}_2(-k)\rangle'}\times \langle {\cal O}_a(k){\cal
O}_b(-k)\rangle'_{\rm L},
\end{equation}
which shows that the power spectra (cosmological correlators
$\langle \phi^p_{{\bf k}}\phi^q_{-{\bf k}}\rangle$) are inversely
proportional to the CFT-correlator and are directly proportional to
the logarithmic part. This is clearly a new observation when one
compares LCFT-correlators with CFT-correlator.

 For a light mass-squared with $m^2 \ll
H^2$, we have $w\simeq \frac{m^2}{3H^2}$. Hence, the corresponding
power spectra are given by
\begin{eqnarray}
&&{\cal P}_{{ab},0}|_{\frac{m^2}{ H^2}\ll1} (k,-\epsilon)=\nonumber
\\
&&\label{ps-light}\xi^2(\epsilon k)^{\frac{2m^2}{3H^2}}\left(
   \begin{array}{ccc}
     0 & 0 &1 \\
     0 & 1 & 2\ln [\epsilon k]+1\\
  1 & 2\ln [\epsilon k]+1 &
   \Big\{2\ln^2[\epsilon k]+\Big(\frac{11}{3}+\frac{4m^2}{27H^2}\Big)\ln[\epsilon k]+\frac{3}{2}\Big\}\end{array}
 \right)
 \end{eqnarray}
 whose spectral indices are given by

\begin{eqnarray}
&&n_{{ab},0}|_{\frac{m^2}{ H^2}\ll1}(k,-\epsilon)-1 =\frac{d\ln{\cal
P}_{{ab},0}|_{\frac{m^2}{ H^2}\ll1}(k,-\epsilon)}{d\ln k}\nonumber\\
\label{sp-light}&&= \left(
\begin{array}{ccc}
0 & 0 & \frac{2m^2}{3H^2} \\
0 & \frac{2m^2}{3H^2} & \frac{2m^2}{3H^2}+\frac{2}{1+2\ln[\epsilon k]}\\
\frac{2m^2}{3H^2} & \frac{2m^2}{3H^2}+\frac{2}{1+2\ln[\epsilon k]} &
\frac{2m^2}{3H^2}+\frac{4\ln[\epsilon
k]+11/3+4m^2/27H^2}{2\ln^2[\epsilon k]+(11/3+4m^2/27H^2)\ln[\epsilon
k]+3/2}\end{array} \right).
\end{eqnarray}
  We observe here that $n_{{ab},0}|_{\frac{m^2}{ H^2}\ll1}$ gets a
 new contribution $\frac{2}{(1+2\ln [\epsilon k])}$ from  the
 logarithmic short distance singularity.

In the massless limit of $m^2=0(\nu=3/2,w=0)$, the corresponding
power spectra take the form
\begin{equation} \label{ps-massless}
{\cal P}_{{ab},0}\Big|_{m^2\to 0}(k,-\epsilon)
=\Big(\frac{H}{2\pi}\Big)^2\left(
   \begin{array}{ccc}
     0 & 0 &1 \\
     0 & 1 & 2\ln [\epsilon k]+1\\
   1 & 2\ln [\epsilon k]+1 &
   2\ln^2[\epsilon k]+\frac{11}{3}\ln[\epsilon k]+\frac{3}{2}\end{array}
 \right)
 \end{equation}
 in the superhorizon limit. This represents the purely log-nature of
 power spectra for a massless three-coupled scalar theory.

\section{Discussions}
We discuss the following issues. \\
$\bullet$ UV and IR boundary conditions in dS inflation \\
In deriving the power spectra of three-coupled scalars, we have
needed two boundary conditions at $\eta=-\infty($UV$,z=\infty)$ and
$\eta=0^-($IR$,z=0)$. The former is necessary to accommodate  the
quantum fluctuations by taking the BD vacuum, while the latter is to
define the LCFT for the dS/LCFT correspondence. These correspond to
the
subhorizon and superhorizon limits, respectively. \\
$\bullet$ Power spectra, LCFT correlators, and the dS/QFT and dS/LCFT correspondences \\
In order to compute the complete power spectrum, we have to solve
the fourth-order and six-order scalar equations on whole dS
spacetime. However, it is formidable to solve these higher-order
equations. Instead, we have obtained two asymptotic solutions at the
UV and IR boundaries. We have gotten   non-trivial commutation
relations (\ref{scft}) which show a feature of the dS/QFT
correspondence in the subhorizon limit. On the other hand, it was
observed from (\ref{p-cc})  that the power spectra in the
superhorizon limit are inversely proportional to the CFT correlator
while they are directly proportional to the logarithmic part. This
shows  that the dS/LCFT correspondence works well in the
superhorizon limit. \\
$\bullet$ Cosmological correlators and LCFT correlators in extrapolate dictionary\\
As was shown in Appendix A, the cosmological correlators in momentum
space
\begin{equation}
\langle\phi^p_{\bf k}\phi^q_{-{\bf k}}\rangle =(2\pi)^4\pi
k^{-3}{\cal P}_{pq}(k)
\end{equation}
are directly proportional to the LCFT correlators $_{\rm e}\langle
{\cal O}_p(k){\cal O}_q(-k)\rangle_{\rm e}$ when one uses the
extrapolate dictionary with operator ${\cal O}_p$ with dimension
$w$ to derive them.  \\
$\bullet$ IR divergence and renormalization \\
To calculate the correlators and power spectra, one has to choose a
proper slice ($\mathbb{R}^3$) near $\eta=0^-$. This has been
performed by taking $\eta=-\epsilon$ firstly, and letting $\epsilon
\to 0$ on later.  Actually, the $\epsilon$-dependence appears in the
power spectra (\ref{ps-matt3}) and spectral indices
(\ref{sp-light}). As was shown in the dS/CFT
correspondence~\cite{Larsen:2003pf}, the cut-off $\epsilon$ acts
like the renormalization scale which is well-known from the UV CFT
renormalization theory.  The cosmic evolution can be seen as a
reversed renormalization group flow, from the IR fixed point (Big
Bang) of the dual CFT to the UV fixed point (Late times) of the dual
CFT theory~\cite{Schalm:2012pi}. Inflation occurs at a certain
intermediate stage during the renormalization group flow as \\
{\it  IR $\longrightarrow$ Inflation $\longrightarrow$ UV} ({\it Big
Bang$\longrightarrow$Inflation$\longrightarrow$ Late times}). \\
This is known to be dS holography. A  choice for $\epsilon$ in dS
spacetime might be the dS scale $H$ and thus, it amounts to
$\epsilon \sim \frac{1}{aH}$. Therefore, in order to obtain the
$\epsilon$-independent power spectra and spectral indices, we must
introduce  proper counter terms to
renormalize the power spectra and spectral indices. \\
$\bullet$ Nonunitarity and truncation \\
As was shown ${\cal P}_{13,0}={\cal P}_{31,0}$ in (\ref{ps-matt3}),
they would be negative for $\ln[\epsilon k]<-1/2$, which implies the
nonunitarity of the power spectrum.  Also, ${\cal P}_{33,0}$ would
be negative for $\frac{6w-11}{2w-3}\ln[\epsilon k]<-2\ln^2[\epsilon
k]-3/2$. These are not acceptable as the power spectra. In order to
address the nonunitarity issue of power spectra, we may propose  to
truncate all log-modes out by imposing appropriate  dS boundary
conditions. After truncation, there will remain a unitary subspace.
This might  be carried out by throwing all modes which generate the
third column and row of the power spectra matrix (\ref{ps-matt3}).
Actually, this is equivalent to throwing all modes which generate
the third column and row of the dual-LCFT matrix (\ref{lcft-mat}).
This is regarded as a truncation process to find a unitary CFT
through the dS/LCFT picture. Hence, the only non-zero power spectrum
is ${\cal P}_{22,0}$ which is surly non-negative. This could be also
proved by using the BRST quantization in the Minkowski spacetime
(equivalently, the truncation process in the dS/QFT correspondence
in the subhorizon
limit)~\cite{Kim:2013waf,Kim:2013mfa}. \\
$\bullet$ Higher-order derivative scalar theory and physical
observables
\\
In this work, we have considered  the three-coupled scalar theory.
We have a second-order equation for $\varphi_1$, a degenerate
fourth-order equation for $\varphi_2$, and a degenerate six-order
equation for $\varphi_3$. Even though $\varphi_2$ is coupled to
 $\varphi_3$ through
(\ref{sing-eq3}), it is a physical field and its power spectrum has
physical relevance. Either the truncation process in the
superhorizon limit  or the BRST quantization in the subhorizon limit
leads  to selecting $\varphi_2$ among $\{\varphi_p\}$. Furthermore,
the three-coupled scalar theory is enough to have a physical power
spectrum.
\\
$\bullet$ Holographic inflation and BICEP2 results \\
Recently, it was shown that  if the dS inflation era of our universe
is approximately described by a dual CFT living on the spatial slice
at the end of inflation (that is, if holographic inflation
occurred), the BICEP2 results might determine the central charge
$c=1.2\times 10^{9}$ of the CFT~\cite{Larsen:2014wpa}. Since the
inflationary era is a dS-like inflation (the slow-roll inflation),
the dual theory must be a near-CFT$_3$. One  can think of it as a
CFT$_3$ perturbed by a nearly marginal operator ${\cal O}$:
$S_u=S_{\rm CFT}+\int d^3x [u {\cal O}]$. In the  single field
inflation, the comoving curvature perturbation $\zeta$ is known to
be conserved at large scales under very general conditions. However,
the authors in~\cite{Garriga:2014ema} has shown that this is not the
case in the dual CFT  description. The requirement that higher
correlators of $\zeta$ should be conserved  restricts the
possibilities for the RG flow.   Imposing such restriction, the
power spectrum $P_\zeta$  must follow an exact power-law. This may
imply that the power-law form of ${\cal P}_{22,0}$ is physically
relevant to the RG flow, even though we did not carry out the
RG-flow in the LCFT.

Consequently, a higher-order derivative scalar theory might  not be
a promising inflation model because it gives rise to the
nonunitarity of power spectra.  Even though the dS/LCFT
correspondence is employed to compute the power spectra, we need to
introduce a truncation process to find a positive (unitary)  power
spectrum for ${\cal P}_{22,0}$.

 \vspace{0.25cm}

 {\bf Acknowledgement}

\vspace{0.25cm}
 This work was supported by the National
Research Foundation of Korea (NRF) grant funded by the Korea
government (MEST) (No.2012-R1A1A2A10040499).

\newpage
\section*{Appendix}
\appendix
\section{LCFT correlators from extrapolate dictionary}

In this appendix, we derive the LCFT correlators by making use of
the extrapolation approach (i) in the superhorizon limit and show
how the relation (\ref{erel}) come out explicitly. For this purpose,
we recall the Green's function for a massive scalar propagating on
dS spacetime \cite{Chernikov:1968zm,Allen:1985wd}
\begin{equation} \label{green}
G_0(\eta,{\bf x};\eta',{\bf
y})=\frac{H^2}{16\pi}\Gamma(\triangle_+)\Gamma(\triangle_-)~_2F_1(\triangle_+,\triangle_-,2;1-\frac{\xi}{4})
\end{equation}
with $\xi=\frac{-(\eta-\eta')^2+|{\bf x}-{\bf y}|^2}{\eta \eta'}$.
Taking  a transformation form of hypergeometric function~\cite{AS}
\begin{eqnarray}
_2F_1(\triangle_+,\triangle_-,2;1-\frac{\xi}{4})=
\Big(\frac{4}{\xi}\Big)^{\triangle_-}~
_2F_1\Big(\triangle_-,2-\triangle_+,2;\frac{1-\frac{\xi}{4}}{-\frac{\xi}{4}}\Big),
\end{eqnarray}
we  obtain the  asymptotic form for $\triangle_-=w$
\begin{equation}\label{g0e}
\lim_{\eta,\eta'\to 0}(\eta\eta')^{-w}G_0(\eta,{\bf x};\eta',{\bf
y})\propto\frac{1}{|{\bf x}-{\bf y}|^{2w}},
\end{equation}
which corresponds to LCFT correlators
\begin{eqnarray}
_{\rm e}\langle {\cal O}_{2}({\bf x}){\cal O}_{2}({\bf
y})\rangle_{\rm e}=_{\rm e}\langle {\cal O}_{1}({\bf x}){\cal
O}_{3}({\bf y})\rangle_{\rm e}=_{\rm e}\langle {\cal O}_{3}({\bf
x}){\cal O}_{1}({\bf y})\rangle_{\rm e}.
\end{eqnarray}
This is the same form as (\ref{oo22}) and  (\ref{oo13}) when
replacing $\triangle_+\to w$.

Furthermore, the Green's functions $G_1$ and $G_2$ are derived by
taking derivative with respect to $w$ as
\begin{eqnarray}
G_1&=&\frac{d}{dw}G_0=\Big(\frac{4}{\xi}\Big)^{w}\Big(-\ln\Big[\frac{\xi}{4}\Big]+\frac{1}{F}\frac{\partial F}{\partial w}\Big)F,\\
&&\nonumber\\
G_2&=&\frac{1}{2}\frac{d}{dw}G_1=\frac{1}{2}\Big(\frac{4}{\xi}\Big)^{w}\Big(\ln^2\Big[\frac{\xi}{4}\Big]-2\ln\Big[\frac{\xi}{4}\Big]\frac{1}{F}\frac{\partial
F}{\partial w}+\frac{1}{F}\frac{\partial^2 F} {\partial w^2}\Big)F,
\end{eqnarray}
where $F$ denotes
$F=H^2\Gamma(3-w)\Gamma(w)_2F_1(w,w-1,2;1-4/\xi)/(16\pi)$. It turns
out that  their asymptotic forms are given by
\begin{eqnarray}
&&\hspace*{-3em}\lim_{\eta,\eta'\to 0}(\eta\eta')^{-w}G_1(\eta,{\bf
x};\eta',{\bf y})\propto\frac{1}{|{\bf x}-{\bf
y}|^{2w}}\Big(-2\ln|{\bf x}-{\bf
y}|+\zeta_1\Big),\label{g1e}\\
&&\hspace*{-3em}\lim_{\eta,\eta'\to 0}(\eta\eta')^{-w}G_2(\eta,{\bf
x};\eta',{\bf y})\propto\frac{1}{|{\bf x}-{\bf
y}|^{2w}}\Big(2\ln^2|{\bf x}-{\bf y}|-2\zeta_1\ln|{\bf x}-{\bf
y}|+\zeta_2\Big),\label{g2e}
\end{eqnarray}
where (\ref{g1e}) and (\ref{g2e}) correspond to
\begin{eqnarray}
_{\rm e}\langle {\cal O}_{2}({\bf x}){\cal O}_{3}({\bf
y})\rangle_{\rm e}=_{\rm e}\langle {\cal O}_{3}({\bf x}){\cal
O}_{2}({\bf y})\rangle_{\rm e}~~~{\rm and}~~~_{\rm e}\langle {\cal
O}_{3}({\bf x}){\cal O}_{3}({\bf y})\rangle_{\rm e},
\end{eqnarray}
being found from (\ref{oo23}) and (\ref{op3op3}), respectively, when
replacing $\triangle_+\to w$. We note that $\zeta_1$ and $\zeta_2$
will be fixed to be finite values after making some regularization
scheme as was shown in (\ref{oo23}) and (\ref{op3op3}).

Finally, we would like to mention that  cosmological correlators
(power spectra) are directly proportional to the LCFT correlators
derived by making use of extrapolate dictionary because
\begin{equation}
\langle\phi^p_{\bf k}\phi^q_{-{\bf k}}\rangle =(2\pi)^4\pi
k^{-3}{\cal P}_{pq}(k) \propto~ _{\rm e}\langle {\cal O}_p(k){\cal
O}_q(-k)\rangle_{\rm e}, \end{equation}
 which is surely compared to the differentiate dictionary in (\ref{p-cc}).

\newpage
\section{LCFT correlators from differentiate  dictionary}

Here, we derive the LCFT correlators by using the differentiation
approach (ii) in the superhorizon limit. In this case, the bulk
bilinear action is given by~\cite{Bergshoeff:2012sc,Kim:2013mfa}
\begin{equation}
\delta S_{\rm S}[\{\varphi_p\}]=-\int_{\rm dS}
d^4x\sqrt{-\bar{g}}\Big[\partial_\mu\varphi_1\partial^\mu\varphi_3+\frac{1}{2}\partial_\mu\varphi_2\partial^\mu\varphi_2
                  +\mu^2\varphi_1\varphi_2
                  +m^2\varphi_1\varphi_3+\frac{1}{2}m^2\varphi^2_2\Big].
\end{equation}
We express the scalar fields $\varphi_a$ in terms of
bulk-to-boundary propagators $K_a$ which relate the bulk solution to
the boundary fields $\varphi_{a,0}$ as
\begin{eqnarray}
\hspace*{-2em}\varphi_1(\eta,{\bf x})&=&\int d^3{\bf y}
\Big[\varphi_{1,0}({\bf y})K_0(\eta,{\bf x};0,{\bf
y})\Big],\label{phi1ss}\label{pbp1}\\
\hspace*{-2em}\varphi_2(\eta,{\bf x})&=&\int d^3{\bf
y}\Big[\varphi_{2,0}({\bf y})K_0(\eta,{\bf x};0,{\bf
y})+\varphi_{1,0}({\bf y})K_1(\eta,{\bf
x};0,{\bf y})\Big]\label{phi2ss},\label{pbp2}\\
\hspace*{-2em}\varphi_3(\eta,{\bf x})&=&\int d^3{\bf
y}\Big[\varphi_{3,0}({\bf y})K_0(\eta,{\bf x};0,{\bf
y})+\varphi_{2,0}({\bf y})K_1(\eta,{\bf x};0,{\bf y})\nonumber \\
\hspace*{8em}&+&\varphi_{1,0}({\bf y})K_2(\eta,{\bf x};0,{\bf
y})\Big].\label{pbp3}
\end{eqnarray}
Here, the bulk-to-boundary propagators $K_a$  satisfy
\begin{eqnarray}
(\bar{\nabla}^2-m^2)K_0&=&0,\label{kk0e}\\
(\bar{\nabla}^2-m^2)K_1&=&\mu^2 K_0,\label{kk1e}\\
(\bar{\nabla}^2-m^2)K_2&=&\mu^2 K_1\label{kk2e},
\end{eqnarray}
and a solution to (\ref{kk0e}) is given by
\begin{eqnarray}\label{gss}
K_0(\eta,{\bf x};0,{\bf y})=c_0\Big[\frac{-\eta}{-\eta^2+|{\bf
x}-{\bf y}|^2}\Big]^{\triangle}.
\end{eqnarray}
Here $c_0$ is a constant and $\triangle$ is determined by
\begin{eqnarray}
\triangle(\triangle-3)H^2+m^2=0,
\end{eqnarray}
whose solution is given by\begin{equation}
\triangle_\pm=\frac{3}{2}\pm\sqrt{\frac{9}{4}-\frac{m^2}{H^2}}.
\end{equation}
We choose $\triangle_+$ only for differentiate dictionary. It is
noteworthy that $K_0$ is not
 a Green's function (bulk-to-bulk propagator) of a massive scalar propagating on dS spacetime.
 Actually, $K_0$ can be derived from the Green's function
 (\ref{green}).
Considering a different transformation for hypergeometric
function~\cite{AS}
\begin{eqnarray}
&&_2F_1(\triangle_+,\triangle_-,2;1-\frac{\xi}{4})=
\Big(\frac{4}{\xi}\Big)^{\triangle_+}~
_2F_1\Big(\triangle_+,2-\triangle_-,2;\frac{1-\frac{\xi}{4}}{-\frac{\xi}{4}}\Big),
\end{eqnarray}
we derive the bulk-to-boundary propagator $K_0$ as
\begin{equation}
\lim_{\eta'\to 0}(\eta')^{-\triangle_+}G_0(\eta,{\bf x};\eta',{\bf
y})\propto\Big[\frac{-\eta}{-\eta^2+|{\bf x}-{\bf
y}|^2}\Big]^{\triangle_+}.
\end{equation}

 Differentiating (\ref{kk0e}) and (\ref{kk1e}) with respect to
$\triangle_+$ and comparing it with (\ref{kk1e}) and (\ref{kk2e})
respectively, the propagators $K_1$ and $K_2$ are found to be
\begin{eqnarray}
K_1(\eta,{\bf x};0,{\bf
y})&=&\frac{d}{d\triangle_+}K_0=K_0\Big(\ln\Big[\frac{-\eta}{-\eta^2+|{\bf
x}-{\bf y}|^2}\Big]+\frac{1}{c_0}\frac{\partial
c_0}{\partial \triangle_+}\Big),\label{g1s}\\
K_2(\eta,{\bf x};0,{\bf
y})&=&\frac{1}{2}\frac{d}{d\triangle_+}K_1~=~\frac{1}{2}\Big(\frac{d}{d\triangle_+}\Big)^2
K_0=\frac{K_0}{2}\Big(\ln^2\Big[\frac{-\eta}{-\eta^2+|{\bf
x}-{\bf y}|^2}\Big]\nonumber\\
&&\hspace*{6.5em}+\frac{2}{c_0}\frac{\partial c_0}{\partial
\triangle_+}\ln\Big[\frac{-\eta}{-\eta^2+|{\bf x}-{\bf
y}|^2}\Big]+\frac{1}{c_0}\frac{\partial^2 c_0}{\partial
\triangle_+^2}\Big),\label{g2s}
\end{eqnarray}
where $\mu^2$ is also determined to be $\mu^2=\partial m^2/\partial
\triangle_+=(3-2\triangle_+)H^2$. Following
~\cite{Ghezelbash:1998rj} where singleton was used to derive the
AdS/LCFT dictionary, we consider $\varphi_{1}(-\epsilon,{\bf x})$
being the Dirichlet boundary value at $\eta=-\epsilon$ near
$\eta=0^-$ and extend it to three-coupled scalar theory.  In this
case, the boundary fields $\varphi_{a,0}({\bf x})$ can be expressed
in terms of $\varphi_{a}(-\epsilon,{\bf x})$
\begin{eqnarray}
\varphi_{1,0}({\bf
x})&\equiv&\epsilon^{\triangle_+-3}\varphi_1(-\epsilon,{\bf
x}),\nonumber\\
\varphi_{2,0}({\bf x})&\equiv&
\epsilon^{\triangle_+-3}\Big[\varphi_2(-\epsilon,{\bf
x})+\ln[\epsilon]\varphi_1(-\epsilon,{\bf x})\Big],\nonumber\\
\varphi_{3,0}({\bf x})&\equiv&
\epsilon^{\triangle_+-3}\Big[\varphi_3(-\epsilon,{\bf
x})+\ln[\epsilon]\varphi_2(-\epsilon,{\bf
x})+\frac{1}{2}\ln^2[\epsilon]\varphi_1(-\epsilon,{\bf
x})\Big].\label{bfs}
\end{eqnarray}
Here we observe asymptotic  behaviors of $\varphi_p(-\epsilon,{\bf
x})$ as $\epsilon \to 0$
\begin{eqnarray} \label{a11}
\varphi_1(-\epsilon,{\bf
x})&\propto&\epsilon^w\varphi_{1,0}({\bf x}),\\
\label{a12}\varphi_2(-\epsilon,{\bf
x})&\propto&\epsilon^w\Big(-\ln[\epsilon]\varphi_{1,0}({\bf x})+\varphi_{2,0}({\bf x})\Big),\\
\label{a13}\varphi_3(-\epsilon,{\bf
x})&\propto&\epsilon^w\Big(\frac{1}{2}\ln^2[\epsilon]\varphi_{1,0}({\bf
x})-\ln[\epsilon]\varphi_{2,0}({\bf x})+\varphi_{3,0}({\bf x})\Big),
\end{eqnarray}
where the first terms in (\ref{a11})-(\ref{a13}) are consistent with
(\ref{su-asym}) for $\eta=-\epsilon$.

Then, we express (\ref{pbp1})-(\ref{pbp3}) as
\begin{eqnarray}
\varphi_1(\eta,{\bf x})&=&c_0\epsilon^{\triangle_+-3}\int d^3{\bf
y}\varphi_1(-\epsilon,{\bf y})\Big[\frac{-\eta}{-\eta^2+|{\bf
x}-{\bf y}|^2}\Big]^{\triangle_+},\\
\varphi_2(\eta,{\bf x})&=&c_0\epsilon^{\triangle_+-3}\int d^3{\bf
y}\Big[\frac{-\eta}{-\eta^2+|{\bf x}-{\bf
y}|^2}\Big]^{\triangle_+}\Big[\varphi_2(-\epsilon,{\bf
y})+\Big(\frac{1}{c_0}\frac{\partial c_0}{\partial
\triangle_+}\nonumber\\
&&\hspace*{13.3em}+\ln\epsilon\Big[\frac{-\eta}{-\eta^2+|{\bf
x}-{\bf y}|^2}\Big]\Big)\varphi_1(-\epsilon,{\bf y})\Big],\\
\varphi_3(\eta,{\bf x})&=&c_0\epsilon^{\triangle_+-3}\int d^3{\bf
y}\Big[\frac{-\eta}{-\eta^2+|{\bf x}-{\bf
y}|^2}\Big]^{\triangle_+}\Big[\varphi_3(-\epsilon,{\bf
y})+\Big(\frac{1}{c_0}\frac{\partial c_0}{\partial
\triangle_+}\nonumber\\
&&\hspace*{-1em}+\ln\epsilon\Big[\frac{-\eta}{-\eta^2+|{\bf x}-{\bf
y}|^2}\Big]\Big)\varphi_2(-\epsilon,{\bf y})+
\Big\{\frac{1}{2}\ln^2\epsilon\Big[\frac{-\eta}{-\eta^2+|{\bf
x}-{\bf y}|^2}\Big]\nonumber\\
&&\hspace*{-0.9em}+\frac{1}{c_0}\frac{\partial c_0}{\partial
\triangle_+}\ln\epsilon\Big[\frac{-\eta}{-\eta^2+|{\bf x}-{\bf
y}|^2}\Big]+\frac{1}{2c_0}\frac{\partial^2 c_0}{\partial
\triangle_+^2}\Big\}\varphi_1(-\epsilon,{\bf y})\Big].
\end{eqnarray}
Now we are in a position to consider an on-shell boundary action
$\delta S_{\rm Sb}$ found from surface integral on the boundary
$\eta=-\epsilon$ after performing some integration by parts
\begin{eqnarray}\label{seff}
\delta S_{\rm Sb}&=&-\frac{1}{2}\lim_{\epsilon\to
0}\int_{\eta=-\epsilon} d^3{\bf
x}\sqrt{\gamma}\left[\varphi_1(\hat{n}\cdot\nabla)\varphi_3
+\varphi_2(\hat{n}\cdot\nabla)\varphi_2+\varphi_3(\hat{n}\cdot\nabla)\varphi_1\right]\nonumber\\
&=&-\frac{1}{2}\lim_{\epsilon\to 0}\int d^3{\bf x}
\epsilon^{-2}\Big[\varphi_1(-\epsilon,{\bf
x})\Big(\frac{\partial\varphi_3(\eta,{\bf
x})}{\partial{\eta}}\Big)_{\eta=-\epsilon}+\varphi_2(-\epsilon,{\bf
x})\Big(\frac{\partial\varphi_2(\eta,{\bf
x})}{\partial{\eta}}\Big)_{\eta=-\epsilon}\nonumber\\
&&\hspace*{12em}+~\varphi_3(-\epsilon,{\bf
x})\Big(\frac{\partial\varphi_1(\eta,{\bf
x})}{\partial{\eta}}\Big)_{\eta=-\epsilon}\Big]\nonumber\\
&=&\frac{1}{2}\lim_{\epsilon\to 0}\int d^3{\bf x}d^3{\bf y}\frac{c_0
\triangle_+}{|{\bf x}-{\bf
y}|^{2\triangle_+}}\epsilon^{2\triangle_+-6}\Big[2\varphi_1(-\epsilon,{\bf
x})\varphi_3(-\epsilon,{\bf y}) + \varphi_2(-\epsilon,{\bf
x})\varphi_2(-\epsilon,{\bf y})\nonumber\\
&&\hspace*{4em}+2\Big(\frac{1}{c_0}\frac{\partial c_0}{\partial
\triangle_+}+\frac{1}{\triangle_+}+2\ln[\epsilon]-2\ln|{\bf x}-{\bf
y}|\Big)\varphi_1(-\epsilon,{\bf x})\varphi_2(-\epsilon,{\bf
y})\nonumber\\
&&\hspace*{4em}+\Big\{\frac{1}{2}(2\ln[\epsilon]-2\ln|{\bf x}-{\bf
y}|)^2+\Big(\frac{1}{\triangle_+}+\frac{1}{c_0}\frac{\partial
c_0}{\partial \triangle_+}\Big)(2\ln[\epsilon]-2\ln|{\bf x}-{\bf
y}|)\nonumber\\
&&\hspace*{10em}+\frac{1}{\triangle_+c_0}\frac{\partial
c_0}{\partial\triangle_+}+\frac{1}{2c_0}\frac{\partial^2
c_0}{\partial \triangle_+^2}\Big\}\varphi_1(-\epsilon,{\bf
x})\varphi_1(-\epsilon,{\bf y})\Big],
\end{eqnarray}
where the normal derivative is defined  by
($\hat{n}\cdot\nabla)=\eta\partial_{\eta}$ and
$\sqrt{\gamma}=1/\eta^3$ with $\gamma$ an induced metric on the
boundary at $\eta=-\epsilon$. Introducing the boundary fields
$\varphi_{a,0}({\bf x})$ [Eq.(\ref{bfs})], we find  the boundary
action (\ref{seff}) which can be written as the classical action
\begin{eqnarray}
&&\delta S_{\rm Sb}[\{\varphi_{a,0}\}]\nonumber\\
&&\hspace*{-3em}=~\frac{1}{2}\int d^3{\bf x}d^3{\bf
y}\frac{c_0\triangle_+}{|{\bf x}-{\bf
y}|^{2\triangle_+}}\Big[2\varphi_{1,0}({\bf x})\varphi_{3,0}({\bf
y})+\varphi_{2,0}({\bf x})\varphi_{2,0}({\bf
y})+2\Big(\frac{1}{\triangle_+}+\frac{1}{c_0}\frac{\partial c_0}{\partial \triangle_+}\nonumber\\
&&-2\ln|{\bf x}-{\bf y}|\Big)\varphi_{1,0}({\bf
x})\varphi_{2,0}({\bf y}) +\Big\{2\ln^2|{\bf x}-{\bf
y}|-2\Big(\frac{1}{\triangle_+}+\frac{1}{c_0}\frac{\partial
c_0}{\partial \triangle_+}\Big) \ln|{\bf x}-{\bf
y}|\nonumber\\
&&\hspace*{12em}+\frac{1}{c_0\triangle_+}\frac{\partial
c_0}{\partial \triangle_+} +\frac{1}{2c_0}\frac{\partial^2
c_0}{\partial \triangle_+^2}\Big\}\varphi_{1,0}({\bf
x})\varphi_{1,0}({\bf y})\Big].\label{sefff}
\end{eqnarray}
Making use of the formula
\begin{equation}
 \langle {\cal O}_{\bar{a}}({\bf x}){\cal O}_{\bar{b}}({\bf
y})\rangle=-\frac{\delta^2\ln Z_{\rm bulk}}{\delta
\varphi_{a,0}({\bf x})\delta \varphi_{b,0}({\bf y})},~~Z_{\rm
bulk}=e^{-\delta S_{\rm Sb}[\{\varphi_{a,0}\}]},
\end{equation}
where $\bar{3}=1,\bar{2}=2, \bar{1}=3$, one can read off the LCFT
correlators  from (\ref{sefff})
\begin{eqnarray}
\label{op1op1}&&\hspace*{-2em} -\frac{\delta^2\ln Z_{\rm
bulk}}{\delta \varphi_{3,0}({\bf x})\delta \varphi_{3,0}({\bf
y})}=\langle {\cal O}_1({\bf x}){\cal O}_1({\bf
y})\rangle=0,\\
\label{oo0}&&\hspace*{-2em} -\frac{\delta^2\ln Z_{\rm bulk}}{\delta
\varphi_{3,0}({\bf x})\delta \varphi_{2,0}({\bf y})}=\langle {\cal
O}_1({\bf x}){\cal O}_2({\bf y})\rangle=\langle {\cal O}_2({\bf
x}){\cal O}_1({\bf
y})\rangle=0,\\
&&\hspace*{-2em} -\frac{\delta^2\ln Z_{\rm bulk}}{\delta
\varphi_{2,0}({\bf x})\delta \varphi_{2,0}({\bf y})}=\langle {\cal
O}_2({\bf x}){\cal O}_2({\bf
y})\rangle~=~\frac{c_0\triangle_+}{|{\bf x}-{\bf
y}|^{2\triangle_+}},\label{oo22}\\
&&\hspace*{-2em} -\frac{\delta^2\ln Z_{\rm bulk}}{\delta
\varphi_{3,0}({\bf x})\delta \varphi_{1,0}({\bf y})}=\langle {\cal
O}_1({\bf x}){\cal O}_3({\bf y})\rangle=\langle {\cal O}_3({\bf
x}){\cal O}_1({\bf y})\rangle~=~\frac{c_0\triangle_+}{|{\bf x}-{\bf
y}|^{2\triangle_+}},\label{oo13}\\
&&\hspace*{-2em}-\frac{\delta^2\ln Z_{\rm bulk}}{\delta
\varphi_{2,0}({\bf x})\delta \varphi_{1,0}({\bf y})}= \langle {\cal
O}_2({\bf x}){\cal O}_3({\bf y})\rangle=\langle {\cal O}_3({\bf
x}){\cal O}_2({\bf
y})\rangle\nonumber\\
&&\hspace*{6.28em}=~\frac{c_0\triangle_+}{|{\bf x}-{\bf
y}|^{2\triangle_+}}\Big(-2\ln|{\bf
x}-{\bf y}|+\frac{1}{\triangle_+}+\frac{1}{c_0}\frac{\partial c_0}{\partial \triangle_+}\Big),
\label{oo23}\\
\label{oo2}&&\hspace*{-2em}-\frac{\delta^2\ln Z_{\rm bulk}}{\delta
\varphi_{1,0}({\bf x})\delta \varphi_{1,0}({\bf y})}=\langle {\cal
O}_3({\bf x}){\cal
O}_3({\bf y})\rangle\nonumber\\
&&\hspace*{6.28em}=~\frac{c_0\triangle_+}{|{\bf x}-{\bf
y}|^{2\triangle_+}}\Bigg(2\ln^2|{\bf x}-{\bf
y}|-2\Big(\frac{1}{\triangle_+}+\frac{1}{c_0}\frac{\partial
c_0}{\partial \triangle_+}\Big) \ln|{\bf x}-{\bf
y}|\nonumber\\
&&\hspace*{18.2em}+\frac{1}{c_0\triangle_+}\frac{\partial
c_0}{\partial \triangle_+} +\frac{1}{2c_0}\frac{\partial^2
c_0}{\partial \triangle_+^2}\Bigg),\label{op3op3}
\end{eqnarray}
which  correspond to the cross coupling given
by~\cite{Bergshoeff:2012sc,Moon:2012vc}
\begin{equation}
\int_{\rm \partial dS_0}d^3{\bf x}\Big[\varphi_{1,0} {\cal
O}_3+\varphi_{2,0} {\cal O}_2+\varphi_{3,0} {\cal O}_1\Big].
\end{equation}

\newpage

\section{Derivation of log-solutions by using the
trick}

It is known that the trick used in~\cite{Kogan:1999bn} indicates how
to solve (\ref{s2-eq2}) directly by differentiating
$(\bar{\nabla}^2-m^2)\varphi_1=0$ with respect to $m^2$. Explicitly,
one can show it by considering the following steps:
\begin{eqnarray}
&&\frac{d}{dm^2}\Big\{\Big(-z^2H^2\frac{d^2}{dz^2}+2z
H^2\frac{d}{dz}-z^2H^2-m^2\Big)\phi_{\bf k}^1(z)
=0\Big\}\\
&\rightarrow&\Big(-z^2H^2\frac{d^2}{dz^2}+2z
H^2\frac{d}{dz}-z^2H^2-m^2\Big)\frac{d}{dm^2}\phi_{\bf
k}^1(z) =\phi_{\bf k}^1(z)\label{phi2to1}\\
&\leftrightarrow&\Big(-z^2H^2\frac{d^2}{dz^2}+2z
H^2\frac{d}{dz}-z^2H^2-m^2\Big)\phi_{\bf k}^2(z) =\mu^2\phi_{\bf
k}^1(z)\label{phi2e1}
\end{eqnarray}
which implies that $\phi_{\bf k}^2(z)$ can be written in terms of
$\phi_{\bf k}^1(z)$ as
\begin{equation} \phi_{\bf k}^2(z)=\mu^2\frac{d}{dm^2}\phi_{\bf
k}^1(z)\label{phi2e}.
\end{equation}
Differentiating (\ref{phi2to1}) further  with respect to $m^2$, one
finds
\begin{eqnarray}
&&\frac{d}{dm^2}\Big\{\Big(-z^2H^2\frac{d^2}{dz^2}+2z
H^2\frac{d}{dz}-z^2H^2-m^2\Big)\frac{d}{dm^2}\phi_{\bf
k}^1(z) =\phi_{\bf k}^1(z)\Big\}\\
&&\rightarrow\Big(-z^2H^2\frac{d^2}{dz^2}+2z
H^2\frac{d}{dz}-z^2H^2-m^2\Big)\Big(\frac{d}{dm^2}\Big)^2\phi_{\bf
k}^1(z) =\frac{2}{\mu^2}\phi_{\bf k}^2(z)\label{91}\\
&&\leftrightarrow\left(-z^2H^2\frac{d^2}{dz^2}+2z
H^2\frac{d}{dz}-z^2H^2-m^2\right)\phi_{\bf k}^3(z) =\mu^2\phi_{\bf
k}^2(z)\label{phi3e1}
\end{eqnarray}
which shows that $\phi_{\bf k}^3(z)$ can be expressed by $\phi_{\bf
k}^1(z)$ as
\begin{equation} \phi_{\bf k}^3(z)=\frac{\mu^4}{2}
\left(\frac{d}{dm^2}\right)^2\phi_{\bf k}^1(z)\label{phi3e}.
\end{equation}
In deriving (\ref{91}),  we have used (\ref{phi2e}). Note that
(\ref{ss2-eq2}) and (\ref{s2-eq2}) can be found by acting
$(\bar{\nabla}^2-m^2)$  on (\ref{phi2e1}) and
$(\bar{\nabla}^2-m^2)^2$ on (\ref{phi3e1}), respectively.
$\frac{d}{dm^2}\phi_{\bf k}^1(z)$ and
$\left(\frac{d}{dm^2}\right)^2\phi_{\bf k}^1(z)$ take the forms
\begin{eqnarray}
\frac{d}{dm^2}\phi_{\bf k}^1(z)&=&-\frac{1}{2\nu
H\sqrt{2k^3}}\sqrt{\frac{\pi}{2}}e^{i\left(\frac{\pi\nu}{2}+\frac{\pi}{4}\right)}z^{3/2}
\Big\{\pi\Big(\frac{i}{2}-\cot[\nu\pi]\Big)H_{\nu}^{(1)}
+i\csc[\nu\pi]\times\nonumber\\
&&\hspace*{10em}\Big(e^{-\nu\pi
i}\frac{\partial}{\partial\nu}J_{\nu}-\frac{\partial}{\partial\nu}J_{-\nu}
-\pi i e^{-\nu\pi i}J_{\nu}\Big)\Big\}\label{phi1e1}
\end{eqnarray}
and
\begin{eqnarray}
\left(\frac{d}{dm^2}\right)^2\phi_{\bf k}^1(z)&=&\frac{1}{4\nu^2
H^3\sqrt{2k^3}}\sqrt{\frac{\pi}{2}}e^{i\left(\frac{\pi\nu}{2}+\frac{\pi}{4}\right)}z^{3/2}
\Big[\Big\{\Big(\frac{\pi}{\nu}-\pi^2i\Big)\cot[\nu\pi]-\frac{5}{4}\pi^2-\frac{\pi}{2\nu}i
\Big\}H_{\nu}^{(1)}
\nonumber\\
&&\hspace*{-5em}-i\csc[\nu\pi]\Big\{\Big(\frac{1}{\nu}+\pi i+2\pi
\cot[\nu\pi]\Big)e^{-\nu\pi
i}\frac{\partial}{\partial\nu}J_{\nu}-\Big(\frac{1}{\nu}-\pi i+2\pi
\cot[\nu\pi]\Big)\frac{\partial}{\partial\nu}J_{-\nu}
\nonumber\\
&&\hspace*{2em}-\Big(\frac{1}{\nu}+2\pi \cot[\nu\pi]\Big)\pi i
e^{-\nu\pi i}J_{\nu}-e^{-\nu\pi
i}\frac{\partial^2}{\partial\nu^2}J_{\nu}
+\frac{\partial^2}{\partial\nu^2}J_{-\nu}\Big\}\Big].\label{phi1e2}
\end{eqnarray}
Here one has to use the relation to find log-solutions as
\begin{eqnarray}
\frac{\partial}{\partial\nu}J_{\nu}(z)=J_{\nu}\ln\Big[\frac{z}{2}\Big]
-\Big(\frac{z}{2}\Big)^{\nu}\sum_{k=0}^{\infty}(-1)^{k}
\frac{\psi(\nu+k+1)}{\Gamma(\nu+k+1)}\frac{(\frac{z^2}{4})^k}{k!}
\end{eqnarray}
with the digamma function $\psi(x)=\partial\ln[\Gamma(x)]/\partial
x$. We observe the appearance of $\ln[z]$-term in (\ref{phi1e1}) and
$\ln^2[z]$-term in (\ref{phi1e2}) when  differentiating the Bessel
function once and twice  with respect to  $\nu$. It turns out that
taking into account $J_{\pm\nu} \to
\Gamma(\pm\nu+1)^{-1}(z/2)^{\pm\nu}$ in the superhorizon limit of
$z\to0$, $\phi_{\bf k}^2(z)$ and $\phi_{\bf k}^3(z)$ take the form
as
\begin{eqnarray} \label{log-phi}
\phi_{\bf k}^2(z)\sim z^{w}\ln[z]~~~{\rm and}~~~\phi_{\bf
k}^3(z)\sim z^{w}\ln^2[z]
\end{eqnarray}
which recover (\ref{super-phi2}) and (\ref{super-phi3}),
respectively.  We point out that
$\frac{\partial}{\partial\nu}J_{-\nu}$ in (\ref{phi1e1}) and
$\frac{\partial^2}{\partial\nu^2}J_{-\nu}$ in (\ref{phi1e2})
contribute to making (\ref{log-phi}) because they behave as
$z^{-\nu}\ln[z]$ and $z^{-\nu}\ln^2[z]$ in the superhorizon limit of
$z\to 0$.  However, it is noted that in the subhorizon limit of
$z\to\infty$, one cannot extract (\ref{z-infinity})  from
(\ref{phi1e1}) and (\ref{phi1e2}) because this trick works in the
superhorizon region only.

\end{document}